\pdfoutput=1
\documentclass[12pt,onecolumn]{IEEEtran}
\usepackage{amsmath,amsfonts,amssymb}
\usepackage{siunitx}
\usepackage{algorithmic}
\usepackage{algorithm}
\usepackage{array}
\usepackage[caption=false,font=normalsize,labelfont=sf,textfont=sf]{subfig}
\usepackage{textcomp}
\usepackage{stfloats}
\usepackage{url}
\usepackage{verbatim}
\usepackage{newunicodechar}
\usepackage{graphicx}
\usepackage{xcolor}
\usepackage{cite}
\usepackage{multirow}
\usepackage{booktabs}
\usepackage{ragged2e}
\usepackage{gensymb}
\usepackage[normalem]{ulem}
\usepackage{upgreek}
\usepackage{enumitem}
\title{A Stochastic Doppler Sensing Framework for Range-Resolved Bidirectional Particle Dynamics in a Bubbling Fluidized Bed at 60 GHz}
\author{Tworit Dash,~\IEEEmembership{Member,~IEEE}, Dingyang Wang,~\IEEEmembership{Member,~IEEE}, Reddy Madhuri Manila, {Wiebren de Jong, and Johan T. Padding}%
\thanks{T. Dash and D. Wang are with the Microwave Sensing, Signals and Systems group, Department of Microelectronics, Delft University of Technology, 2628 CD Delft, The Netherlands {(e-mail: t.k.dash@tudelft.nl).}}%
\thanks{{R.M. Manila, W. de Jong, and J. T. Padding} are with the Department of Process and Energy, Delft University of Technology, 2628 CB Delft, The Netherlands {(e-mail: J.T.Padding@tudelft.nl).}}}
\begin{document}
\maketitle
\begin{abstract}
The problem of observing particle motion inside an optically opaque, dense bubbling fluidized bed is addressed. Differential pressure measurements identify the bulk fluidization state but do not reveal where particles move or how their velocities vary with height. A commercial frequency-modulated continuous-wave (FMCW) radar near \SI{60}{GHz} is mounted above the facility to obtain depth-resolved measurements without optical access or an intrusive probe. {The measured depth interval is divided into \SI{4.34}{cm} cells, each of which contains an ensemble of \SI{600}{\micro m} sand grains. At \SI{60}{GHz}, these particles fall within the Rayleigh--Mie scattering transition region.} Particle displacement, cell exchange, and changes in shadowing alter the phases and amplitudes of the echoes. Their coherent sum is therefore a stochastic complex signal, while its covariance and Doppler spectrum describe the ensemble motion. A novel statistical spectral analysis retrieves total power, mean line-of-sight (radial) velocity, and Doppler spectral width over range and time. Eight-minute measurements reveal an expanding radar-detectable layer and alternating motion towards and away from the radar. Positive- and negative-mean regions coexist during developed fluidization and exchange their contribution to the particle power. Most detectable spectra contain one dominant Doppler lobe; a small subset supports two distinguishable lobes, {including cases with one negative and one positive mean velocity}. {The 5-s total-power trend has a correlation of $0.734$ with the measured pressure drop, whereas the local fitted spectral-width trend has a weaker correlation of $0.152$.} The novelty is a range-resolved statistical Doppler measurement that reveals bidirectional particle motion and determines when multiple local velocity contributions are distinguishable. {To the authors' knowledge, these are the first such Doppler measurements with a commercial Texas Instruments \SI{60}{GHz} radar.} The statistical structure of the Doppler spectrum of these complex random radar signals, and its evolution in space and time in dense bubbling fluidized beds, appear to constitute a previously unexplored domain.
\end{abstract}

\section{Introduction}\label{sec:introduction}
Renewable energy supply combined with circular carbon processing is central to reducing anthropogenic carbon-dioxide emissions and replacing fossil feedstocks. Gasification provides a versatile thermochemical route for converting low-cost and low-grade carbonaceous waste into synthesis gas, whose principal combustible constituents are carbon monoxide and hydrogen. Fluidized-bed reactors are attractive for this conversion because they accommodate broad variations in feedstock composition, particle size, and shape while providing effective gas-solid contact, mixing, and heat and mass transfer. These properties support high reaction efficiency and make fluidized beds suitable for scalable gasification and other large industrial processes \cite{Kunii}.

Extensive practical experience is available for fluidized-bed conversion of fossil fuels and several biomass feedstocks, including waste wood. More heterogeneous streams, such as sewage-derived material and mixed plastics, remain difficult to convert into clean synthesis gas. The hydrodynamics of gas-solid fluidization must therefore be characterized accurately to improve reactor performance {and improve operating stability and scale-up}. Relevant quantities include bubble formation and eruption, particle mixing, solids circulation, pressure fluctuations, bed expansion, concentration, and particle-velocity distributions.

Differential pressure and expanded-bed height provide the most common indicators of the fluidization state. More detailed local information can be obtained with high-speed imaging, particle-image velocimetry, and particle-tracking velocimetry, but optical methods are generally restricted to transparent or pseudo-two-dimensional beds \cite{xinPIVPTv,FULLMER2020323}. Electrical-capacitance tomography is non-intrusive, yet its implementation is costly, its spatial resolution is limited, and image formation requires an inverse reconstruction \cite{saied2016,Huang}. X-ray tomography \cite{chen2024x,van2016fast,escudero2016characterizing}, gamma-ray computed tomography \cite{MacCuaig.N,Abdelsalam.E}, and magnetic-resonance imaging \cite{rees2006nature,muller2009geometrical} can penetrate opaque beds, but safety requirements, installation volume, and acquisition complexity limit their routine use. Reviews of hydrodynamic measurement methods for gas-solid fluidized beds are given in \cite{Werther,Zhu}.

Radar provides depth and motion information without optical access or instrumentation inside the bed. {Although industrial radar is commonly used for level measurement, range-resolved Doppler sensing of dense multiphase systems remains much less established.} Mounted above the vessel, the radar resolves depth cells both within the dense bed and in the freeboard zone and measures the particle-velocity component along its line of sight. Because the line of sight is approximately vertical, this radial component corresponds primarily to the axial particle velocity in the bed. It therefore complements {pressure measurements}: pressure records the bulk mechanical response produced by gas-solid momentum transfer and bed restructuring, whereas radar indicates where the scattering particles move within that bulk state. Experiments at {Chalmers University of Technology} have demonstrated {the usefulness of} submillimetre-wave measurements near \SI{340}{GHz} in circulating fluidized beds and particle clouds, principally in leaner or dilute regions \cite{Bonmann2022,GuioPerez2023,Bryllert2023}. Their large bandwidth gives approximately centimetre-scale range resolution and a small measurement volume. {The radar used in this work} operates near \SI{60}{GHz}. Its available bandwidth gives coarser range resolution, but the longer wavelength reduces the electrical size of the sand particles and offers a different penetration-resolution compromise {for} dense, solids-rich bed{s}.

{Range--Doppler spectra describe} how the received power is distributed over depth and radial velocity. Each range cell contains many electrically small particles, so the radar does not resolve their individual trajectories. Instead, it measures how the returned power is distributed over radial velocity. The first three Doppler moments summarize this distribution in physically meaningful quantities: total power describes the strength and spatial extent of the particle return; mean Doppler describes its net radial direction; and spectral width describes the spread of the contributing radial velocities. A mean close to zero does not imply an immobile bed, because returns from particles moving towards and away from the radar can balance. Total power (zeroth Doppler moment), mean Doppler (first Doppler moment), and width (square root of the second central Doppler moment) must therefore be interpreted together.

For a central wavelength $\lambda$ and radial displacement $\delta R$, the phase of a monostatic radar echo changes by $4\pi\delta R/\lambda$. A displacement of only $\lambda/2$ therefore produces a complete phase cycle. At the present wavelength of \SI{4.834}{mm}, millimetric rearrangements substantially change the coherent sum. The exact complex sample cannot be predicted because the particle positions, velocities, propagation paths, and membership of a range cell are not known grain by grain. The slow-time return is therefore treated as a random process: its individual samples fluctuate, but their covariance over sweep lag and the corresponding Doppler spectrum have repeatable structure. Randomness does not mean that the return is only receiver noise; the statistical structure is produced by the moving particle ensemble.

Classical mean-frequency (corresponding to the mean Doppler velocity) estimators provide an important reference for Doppler processing \cite{Sirmans1975}. Parametric covariance and spectral estimators developed for distributed radar targets further show how total power, mean Doppler, and width can be retrieved from a stochastic return \cite{Dash2023PrecipitationPrior,Dash2024DopplerModel,Dash2024CounterAliasing,Dash2024Incoherent}. The present analysis first separates a fitted extended Doppler-lobe contribution from the background measured before fluidization. One extended Doppler lobe then describes the dominant local ensemble motion. Unlike conventional point-target Doppler measurements, an extended Doppler lobe here represents the combined return from many particles whose radial velocities are distributed about a common mean. Its width describes the velocity dispersion of that scattering ensemble. Additional extended Doppler lobes can occur within the same resolution cell when distinguishable particle ensembles move with different central velocities. They are introduced only where the measured spectral shape cannot be represented adequately by a single extended Doppler lobe. The Akaike information criterion (AIC) and Bayesian information criterion (BIC) compare {such an improvement} with the extra number of fitted parameters. {The order of this procedure} preserves the physical question: first, whether particle motion is measurable; second, whether more than one extended Doppler lobe is distinguishable in the same range-time cell.

The novelty and contributions {of this work} are summarized as follows:
\begin{itemize}[leftmargin=*,nosep]
\item range-resolved Doppler measurements inside a dense bubbling fluidized bed using a compact commercial Texas Instruments radar at \SI{60}{GHz};
\item a measured-background spectral formulation that retrieves total power, mean radial velocity, and Doppler spectral width, together with a background-calibrated measure of Doppler-spectrum detectability;
\item complete-record and focused range-time observations of alternating positive and negative mean Doppler, together with power-weighted directional quantities that expose irregular motion during developed fluidization;
\item a hierarchical AIC and BIC assessment that determines when one effective extended Doppler lobe is sufficient and when two or three spectral contributions are supported.
\end{itemize}
The physical observation of bidirectional motion is established from the one-lobe range-resolved retrieval before additional extended Doppler lobes are considered. The multi-lobe analysis then tests whether opposite velocity contributions are simultaneously distinguishable within an individual range-time cell.

The main body of the paper is organized as follows. Section~\ref{sec:experiment} describes the fluidized-bed facility, radar measurement, and scattering regime. Section~\ref{sec:signal_model} formulates the stochastic slow-time return, range-Doppler processing, Doppler-spectrum detectability criterion, and multi-lobe estimator. Section~\ref{sec:results} presents the range-Doppler observations, one-lobe Doppler retrievals, {comparison with pressure measurements}, bidirectional particle motion, and the subsequent AIC and BIC assessment of spectral complexity. Section~\ref{sec:discussion} discusses the physical interpretation and resolution limitations. Section~\ref{sec:conclusion} concludes the paper.

\section{Experimental Arrangement}\label{sec:experiment}
\subsection{Fluidized-bed facility}
\begin{figure}[t]
\centering
\includegraphics[width=0.50\textwidth]{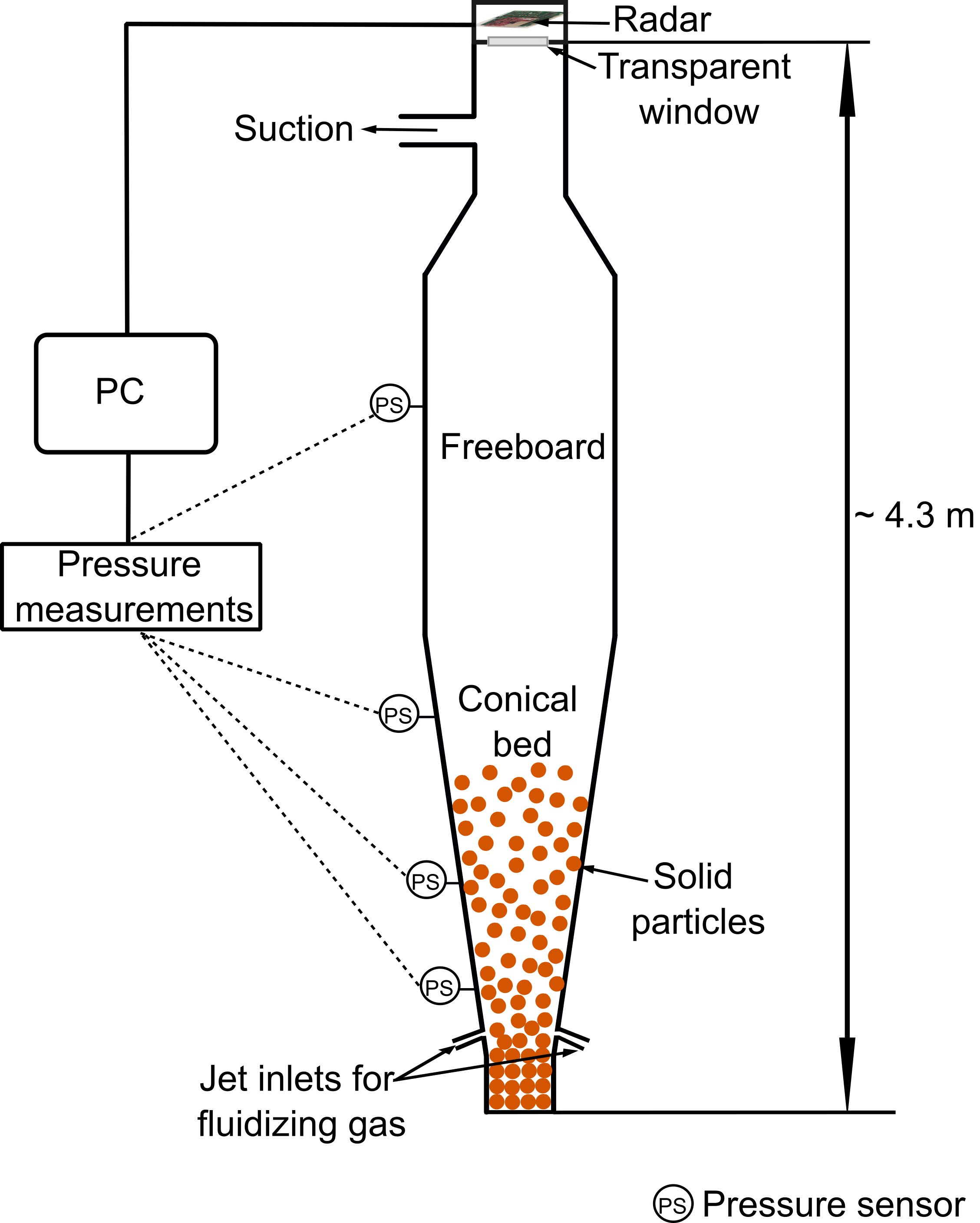}
\caption{Conical jet-blown bubbling fluidized bed with the \SI{60}{GHz} radar mounted at the top of the fluidized-bed facility, above the vessel. The indicated \SI{4.3}{m} dimension extends from the transparent window to the vessel bottom; radar range is referenced to the phase centre above the window. Nitrogen enters through three inclined wall nozzles, and wall-mounted pressure sensors record the mechanical bed response.}
\label{fig:setup}
\end{figure}

Fig.~\ref{fig:setup} shows the conical jet-blown bubbling fluidized bed, which is made of stainless steel (S235). Nitrogen is used as the fluidizing gas and is supplied through three nozzles positioned equidistantly around the conical wall. The nitrogen flow contains no appreciable water-vapor load, so variable attenuation by moisture in the process gas is not expected to affect the radar measurement. The nozzles are located \SI{50}{mm} above the cone base and are inclined by \SI{20}{\degree} with respect to the horizontal. The bed material is silica sand with a mean particle diameter of approximately \SI{600}{\micro m} and a particle density of approximately \SI{2730}{kg.m^{-3}}. Pressure sensors at different axial heights provide the bed-pressure signal used to identify the minimum-fluidization condition. {Signals from the pressure-sensor system are acquired at \SI{50}{Hz}.} The mass-flow controllers and pressure sensors are integrated with National Instruments LabVIEW for control and acquisition.

The radar is mounted above the facility and observes vertically. {A Perspex window that is transparent to the radar signal protects the radar from fines generated during fluidization.} The window-to-bottom distance is approximately \SI{4.3}{m}, while the radar phase center lies above the window. The direct bottom response is consequently expected slightly beyond \SI{4.3}{m} on an axis referenced to the radar. The quantitative analysis is restricted to the apparent-range interval \SIrange{2.518}{4.489}{m}. This upper limit includes the direct bottom response and its range-resolution main lobe while excluding the delayed stationary response observed near \SI{4.665}{m}. Returns at larger apparent range can arise from propagation through the dense bed and reflections from the metallic vessel; they are not assigned to geometric positions below the vessel bottom. The measurements are performed at ambient conditions. The gas flow {through} each nozzle is varied stepwise from \SIrange{50}{350}{L.min^{-1}} to traverse fixed-bed, minimum-fluidization, and fully fluidized conditions. The analysis uses the first \SI{8}{min} of the record, which contains the complete imposed flow sequence. Radar and pressure are placed on their recorded time coordinates, and the pressure trace is interpolated to the radar times without an additional time-lag correction.

\subsection{Frequency-modulated continuous-wave Doppler measurement}
A frequency-modulated continuous-wave (FMCW) radar transmits repeated linear frequency sweeps. After dechirping, the beat frequency of a target at range $R$ and radial velocity $v$ is approximately
\begin{equation}
 f_b\simeq\frac{2\gamma R}{c}+\frac{2v}{\lambda},
 \label{eq:beat_frequency}
\end{equation}
where $\gamma$ is the sweep slope, $c$ is the speed of light, and $\lambda$ is the wavelength. Even at the unambiguous velocity limit, the Doppler term corresponds to less than \SI{1.8}{cm} of apparent range displacement, below the \SI{4.34}{cm} range resolution. The range-dependent term therefore dominates the beat frequency. The delay-derived quantity is the equivalent free-space, or apparent, range
\begin{equation}
 \widetilde R\simeq\frac{cf_b}{2\gamma}.
 \label{eq:range}
\end{equation}
For a direct path in air, $\widetilde R$ equals the geometric distance from the radar phase centre. Propagation through dielectric material or along a reflected path increases delay, so a return can appear at $\widetilde R$ larger than the position of its scatterer. This distinction is important near the metallic vessel bottom. Velocity is obtained from the phase progression between sweeps. With $N_s$ acquired fast-time samples at sampling rate $f_s$, the processed bandwidth is $B=\gamma N_s/f_s$, and the corresponding range resolution is
\begin{equation}
 \Delta R=\frac{c}{2B}.
 \label{eq:range_resolution}
\end{equation}
{Thus, the beat-frequency variation within a sweep determines apparent range, whereas the phase progression across repeated sweeps determines radial velocity within each range cell.}

The radar front end is a Texas Instruments IWR6843 single-chip \SIrange{60}{64}{GHz} sensor, and the raw digitized receiver samples are recorded with a Texas Instruments DCA1000 evaluation-module data-capture card \cite{TI6843,TIDCA1000}. The acquisition configuration activates one transmitter and four receivers. Each receiver is processed separately through the range and Doppler transforms, after which the four spectra are averaged in power. 


The IWR6843ISK module has a nominal field of view of approximately $120^\circ$ in azimuth and $30^\circ$ in elevation \cite{TI6843ISK}. These values describe the module field of view. For a field-of-view angle $\Theta_\psi$ in plane $\psi\in\{\mathrm{az},\mathrm{el}\}$, the corresponding free-space cross-range span at range $R$ is
\begin{equation}
 W_\psi(R)=2R\tan\!\left(\frac{\Theta_\psi}{2}\right).
 \label{eq:footprint}
\end{equation}
At \SI{5}{m}, the nominal azimuth and elevation spans are approximately \SI{17.3}{m} and \SI{2.68}{m}, respectively, both larger than the vessel cross-section. The illuminated bed volume is therefore set primarily by the intersection of the broad antenna response, the range shell, and the vessel. {Finer angular separation requires coherent processing of the receive array and will be the topic of future work.}

\begin{table}[t]
\caption{Radar waveform, measurement, and spectral-retrieval parameters.}
\label{tab:parameters}
\centering
\begin{tabular}{@{}>{\RaggedRight\arraybackslash}p{0.24\textwidth}>{\RaggedRight\arraybackslash}p{0.21\textwidth}>{\RaggedRight\arraybackslash}p{0.27\textwidth}>{\RaggedRight\arraybackslash}p{0.20\textwidth}@{}}
\toprule
Radar or waveform quantity & Value & Processing quantity & Value \\
\midrule
Radar hardware & IWR6843/DCA1000EVM & Fast-time samples per sweep & 230 \\
Sweep start/effective centre frequency & \SI{60}{GHz}/\SI{62.018}{GHz} & Range-transform length & 512 \\
Wavelength & \SI{4.834}{mm} & Slow-time samples per spectrum & 128 \\
Frequency slope & \SI{60}{MHz/\micro s} & Doppler transform, retrieval/visual & 128/256 \\
Processed bandwidth & \SI{3.450}{GHz} & Recorded update interval & \SI{0.05}{s} \\
Range resolution & \SI{4.34}{cm} & Retrieval/visual velocity spacing & \SI{0.2671}{m/s}/\SI{0.1336}{m/s} \\
Fast-time sampling rate & \SI{4.000}{MS/s} & Independent Doppler-bin spacing & \SI{0.2671}{m/s} \\
Sweep-repetition interval & \SI{70.69}{\micro s} & Coherent processing interval & \SI{9.05}{ms} \\
Sampling start/ramp-end/idle times & \SI{4.88}{\micro s}/\SI{63.69}{\micro s}/\SI{7}{\micro s} & Frame interval & \SI{50}{ms} \\
Transmit/receive channels & 1/4 & Unambiguous velocity magnitude & \SI{17.096}{m/s} \\
Sweeps per recorded frame & 255 & Sweeps used per Doppler record & 128 \\
Recorded/analyzed frames & 12000/9600 & Evaluated range cells & 102 \\
Nominal field of view, azimuth/elevation & $120^\circ/30^\circ$ & Evaluated apparent-range interval & \SIrange{2.518}{4.489}{m} \\
Angular output & one power-averaged channel & Fitted non-oversampled ordinates & 128 \\
Background interval & \SIrange{6}{27}{s} & Detectability decision & calibrated extended-lobe/background contrast \\
\bottomrule
\end{tabular}
\end{table}

\subsection{Scattering by a particle ensemble}
At \SI{62.018}{GHz}, the wavelength is approximately \SI{4.834}{mm}. The electromagnetic size parameter for the mean particle diameter is
\begin{equation}
 \xi=\frac{\pi d_p}{\lambda}\approx0.39.
 \label{eq:size_parameter}
\end{equation}
The particles are electrically small and lie on the Rayleigh side of the Rayleigh-Mie transition rather than in the geometric-optics regime \cite{BohrenHuffman1983}. The dense bed nevertheless cannot be reduced to isolated ideal Rayleigh scatterers: particle-size dispersion, dielectric properties, shadowing, attenuation, and multiple scattering all influence the received power. Absolute power is therefore not converted directly into solids concentration.

The radar observes an ensemble rather than individual grains. For a range cell centered at $R$, let $A_b(R)$ denote the vessel cross-sectional area intercepted by the antenna response and let $V_b(R)\simeq A_b(R)\Delta R$ denote the corresponding bed volume. Let $\phi_s$ denote the local solids fraction. For monodisperse spherical grains with volume $\pi d_p^3/6$, the approximate mean number of grains in that bed volume is

\begin{equation}
 \overline M(R)\approx
 \frac{6\phi_sV_b(R)}{\pi d_p^3}.
 \label{eq:particle_number}
\end{equation}

This is a geometric population estimate, not a radar retrieval of particle count. Only one litre of bed at $\phi_s=0.5$ contains approximately $4.4\times10^6$ grains of this diameter. The complex fields from these grains and their propagation paths add coherently at the receiver. Their amplitudes are altered by attenuation and shadowing, while their phases change with radial displacement. Grains also enter and leave a range shell as the bed circulates. The microscopic configuration, therefore, differs from one sweep to the next (a sweep corresponds to one full chirp time), even when pressure and gas flow are nearly constant.

This physical variability produces a random process indexed by slow time. {This does not imply independent or structureless samples.} Successive echoes remain correlated while the ensemble retains memory of its earlier configuration. The rate of phase {change} produces the mean Doppler frequency, and the loss of correlation with sweep separation reflects the spread of radial velocities and the changing membership of the cell. The Doppler spectrum is the frequency-domain representation of this covariance. It consequently describes the statistics of the ensemble motion, rather than the trajectory of a particular grain.

\section{Doppler Signal Model and Spectral Representation}\label{sec:signal_model}
\subsection{Stochastic particle return and measured background}
The notation follows the radar processing sequence from the measured analogue-to-digital-converter sample to the fitted Doppler spectrum. Let $x_{q,t}[p,n]$ denote the complex dechirped sample from receive channel $q$ in recorded frame $t$, where $t$ labels both the frame and its recorded time coordinate, $p=0,\ldots,N_s-1$ is the fast-time sample within sweep $n$, and $n=0,\ldots,N-1$ is the sweep index within the coherent interval. After applying the fast-time Hann window $w_R[p]$, the range-transform output at integer range bin $m$ is
\begin{equation}
z_{q,m,t}[n]=\sum_{p=0}^{N_s-1}x_{q,t}[p,n]w_R[p]
\exp\!\left(-j\frac{2\pi mp}{N_R}\right),
\label{eq:range_fft}
\end{equation}
where $N_R$ is the range-transform length. For sampling frequency $f_s$, chirp slope $\gamma$, and propagation speed $c$, bin $m$ corresponds to the calibrated range
\begin{equation}
r_m=\frac{cf_s}{2\gamma N_R}m.
\label{eq:range_axis}
\end{equation}
The remainder of the analysis uses the physical range coordinate $r=r_m$ rather than the integer bin $m$. The range-resolved slow-time sample is therefore defined by
\begin{equation}
y_{q,r_m,t}[n]\equiv z_{q,m,t}[n].
\label{eq:range_to_slow_time}
\end{equation}
Thus, $x$ is the measured fast-time sample, $z$ is its range-transform output, and $y$ is the same range-resolved quantity indexed by calibrated physical range. For receive channel $q$, range cell $r$, frame $t$, and sweep $n$, this slow-time return is decomposed as
\begin{align}
y_{q,r,t}[n]
&=s_{q,r,t}[n]+b_{q,r,t}[n],\nonumber\\
s_{q,r,t}[n]
&=\sum_{i\in\mathcal I_{r,t}[n]}
A_{q,r,i,t}[n]\nonumber\\
&\quad\times
\exp\!\left(
j\phi_{q,r,i,t}[n]
+j\frac{4\pi R_{i,t}[n]}{\lambda}
\right).
\label{eq:field_sum}
\end{align}
Here, $s_{q,r,t}[n]$ is the moving-particle contribution and $b_{q,r,t}[n]$ is the measured background contribution; both are components of the measured range-resolved sample $y_{q,r,t}[n]$, not separately observed signals. The symbol $j$ is the imaginary unit. {The set $\mathcal I_{r,t}[n]$ is the summation set in \eqref{eq:field_sum}: it contains the particle-scattering and propagation-path contributions present in range cell $r$ during sweep $n$. Its cardinality, $M_{r,t}[n]=|\mathcal I_{r,t}[n]|$, is the instantaneous number of contributions in that coherent sum. For contribution $i$, $A_{q,r,i,t}[n]$ is its received amplitude, $\phi_{q,r,i,t}[n]$ is its residual scattering and propagation phase, and $R_{i,t}[n]$ is its radar range.} The background term contains the stationary fixed-bed and vessel returns, coherent leakage, receiver contribution, and persistent propagation paths. It is measured during the no-motion interval and is retained in the Doppler spectrum, including its strong contribution near zero radial velocity.

Particle entry and exit change $\mathcal I_{r,t}[n]$, while particle motion and changing propagation paths change the phases in \eqref{eq:field_sum}. The particle return is therefore the coherent sum of many weak, differently phased fields. When no single contribution dominates, the central-limit argument gives approximately Gaussian in-phase and quadrature components and hence a proper complex Gaussian particle process \cite{Goodman2015,Dash2023PrecipitationPrior,Dash2024DopplerModel}. This convergence concerns the received field and does not imply that the particle velocities themselves are Gaussian. The background is described by its measured second-order structure rather than by assigning its zero-Doppler power to particle motion.

Over one coherent observation interval, the particle and background statistics are assumed to vary slowly enough that their second-order quantities depend mainly on sweep separation. This is the local-stationarity approximation. Stacking the $N$ range-resolved samples gives
\begin{equation}
\mathbf y_{q,r,t}
=
\begin{bmatrix}
y_{q,r,t}[0]&\cdots&y_{q,r,t}[N-1]
\end{bmatrix}^{T}.
\label{eq:slow_time_vector}
\end{equation}
The vector is formed directly from the $N$ values of $y_{q,r,t}[n]$ defined in \eqref{eq:range_to_slow_time}; no empirical slow-time mean is subtracted. Assuming that the particle and background terms are mutually uncorrelated over this interval, and that the calibrated receiver channels share the same second-order model, the working proper-complex Gaussian model for each channel is

\begin{align}\label{eq:data_model}
\mathbf y_{q,r,t}
&\sim
\mathcal{CN}\!\left(
\mathbf 0,\mathbf C_{r,t}
\right),\nonumber\\ 
\mathbf C_{r,t}
&=
\mathbf C_{s,r,t}(\boldsymbol{\theta}_{r,t})
+a_{r,t}\mathbf C_{b,r},\\ \nonumber
\boldsymbol{\theta}_{r,t}
&=
\left\{
\widetilde\mu_{r,t},\widetilde\sigma_{r,t},P_{r,t}
\right\}.
\end{align}

The zero vector in \eqref{eq:data_model} is the ensemble-mean convention of the second-order random-field model; it does not represent subtraction of the mean of each measured slow-time record. The retained coherent power, including the central zero-Doppler response, is represented by the measured background covariance $\mathbf C_{b,r}$ and its spectral counterpart introduced below. The positive factor $a_{r,t}$ adapts the background power to each range--time cell while preserving its measured spectral shape.

In \eqref{eq:data_model}, $\mathcal{CN}$ denotes a proper complex Gaussian distribution, $\widetilde\mu_{r,t}$ is the dimensionless mean Doppler frequency in cycles per sweep, $\widetilde\sigma_{r,t}$ is its dimensionless spectral width, $P_{r,t}$ is the extended-Doppler-lobe power, and $\mathbf C_{s,r,t}$ is its covariance matrix. The tilde distinguishes normalized frequency quantities from their physical-unit velocity counterparts introduced below. Constant transform gains are absorbed into $P_{r,t}$ and the background term, so the reported power is uncalibrated but comparable over range and time. The covariance sequence and covariance matrix are related by
\begin{align}
R_{y,r,t}[h]
&=\mathbb E\!\left\{
y_{q,r,t}[n+h]y_{q,r,t}^{*}[n]
\right\},\\
\left[\mathbf C_{r,t}\right]_{a,b}
&=R_{y,r,t}[a-b],
\qquad a,b=0,\ldots,N-1,
\label{eq:covariance_sequence_matrix}
\end{align}
where {$\mathbb E\{\cdot\}$ denotes statistical expectation, $h$ is the integer sweep lag, and $(\cdot)^*$ denotes the complex conjugate}. The expectation is over possible realizations of the random particle configuration at the same operating condition; it is not an average over the entries of the measured vector. Thus, $R_{y,r,t}[h]$ is one scalar covariance value at lag $h$, whereas $\mathbf C_{r,t}$ collects these values into the finite-sample covariance matrix. Indices $a$ and $b$ identify rows and columns of this matrix and are distinct from the Doppler-bin index $u$ introduced below. {For one Gaussian-shaped extended Doppler lobe, the particle and total covariance sequences are}
\begin{align}
R_{s,r,t}[h]
&=P_{r,t}
\exp\!\left(j2\pi h\widetilde\mu_{r,t}\right)
\exp\!\left(-2\pi^2h^2\widetilde\sigma_{r,t}^{2}\right),\\
R_{y,r,t}[h]
&=R_{s,r,t}[h]+a_{r,t}R_{b,r}[h],
\label{eq:gaussian_cov}
\end{align}
where $R_{b,r}[h]$ is the background covariance sequence represented by $\mathbf C_{b,r}$. The rotating factor in $R_{s,r,t}[h]$ describes the average phase progression and hence the mean radial motion. Its decaying magnitude describes the loss of coherence with sweep lag; larger $\widetilde\sigma_{r,t}$ gives faster decay and a broader radial-velocity distribution. Equations~\eqref{eq:field_sum}--\eqref{eq:gaussian_cov} define the background-aware stochastic model underlying the spectral estimator \cite{Frehlich1993,Dash2024DopplerModel}.
If $f_D$ is {the} Doppler frequency, $v$ is {the} radial velocity, and $\lambda$ is {the} wavelength, the normalized Doppler frequency is
\begin{equation}
\widetilde f_D=f_DT_s=\frac{2vT_s}{\lambda},\label{eq:normalization}
\end{equation}
where $T_s$ is the sweep-repetition interval of the radar waveform.
Thus, the normalized Doppler frequency $\widetilde f_D$ is measured in cycles per sweep and lies in $[-0.5,0.5)$. Its mean $\widetilde\mu$ and width $\widetilde\sigma$ convert to physical velocity units according to
\begin{equation}\mu_v=\frac{\lambda}{2T_s}\widetilde\mu,\qquad \sigma_v=\frac{\lambda}{2T_s}\widetilde\sigma.\label{eq:conversion}
\end{equation}
With the phase convention in \eqref{eq:field_sum}, positive velocity denotes increasing range (motion away from the radar), whereas negative velocity denotes decreasing range (motion towards it). For an $N_D$-point Doppler spectrum, the radial-velocity coordinate is
\begin{equation}
v_u=\frac{\lambda}{2T_s}\left(\frac{u-N_D/2}{N_D}\right),\qquad u=0,\ldots,N_D-1.
\label{eq:velocity_axis}
\end{equation}
For the {measurements presented in this work}, $T_s=\SI{70.69}{\micro s}$ and the effective centre frequency is \SI{62.018}{GHz}. Parameter estimation uses $N_D=N=128$, giving a velocity spacing and independent velocity resolution of \SI{0.2671}{m/s} over the unambiguous interval of approximately $\pm\SI{17.096}{m/s}$. A 256-point transform, with \SI{0.1336}{m/s} displayed spacing, is used only for the range-Doppler illustrations and does not enter the likelihood or the information criteria.

\subsection{Range-Doppler spectrum formation}
Range-Doppler processing separates two physical coordinates. Equations~\eqref{eq:range_fft} and \eqref{eq:range_axis} use the frequency variation within each transmitted sweep to locate the scattering volume in depth. The following slow-time transform uses the sweep-to-sweep phase variation of $y_{q,r,t}[n]$ to describe radial motion inside that range cell. Here, $N_s$ and $N$ are the fast- and slow-time sample counts, while $N_R$ and $N_D$ are the corresponding transform lengths. Each frame is repeated every \SI{50}{ms}. The coherent portion used for Doppler estimation contains $N=128$ sweeps and spans \SI{9.05}{ms}; one retrieval is obtained at each frame time throughout the eight-minute interval. The range-resolved samples enter the slow-time transform directly, so the stationary zero-Doppler contribution remains available to the measured background model.

Parameter estimation uses a rectangular slow-time record and the non-oversampled $N_D=N=128$ point discrete Fourier transform (DFT). Its coefficient at centered Doppler-bin index $u$ is
\begin{equation}
d_{q,r,t}[u]=\sum_{n=0}^{N-1}y_{q,r,t}[n]
\exp\!\left[-j2\pi\left(\frac{u-N/2}{N}\right)n\right],
\label{eq:doppler_fft}
\end{equation}
where the shift by $N/2$ centers zero Doppler. The discrete normalized Doppler frequency associated with index $u=0,\ldots,N-1$ is
\begin{equation}
\widetilde f_u=\frac{u-N/2}{N},
\qquad -\frac{1}{2}\leq \widetilde f_u<\frac{1}{2}.
\label{eq:centered_frequency_axis}
\end{equation}
The corresponding radial velocity is obtained from \eqref{eq:velocity_axis}. The measured receiver-averaged periodogram is
\begin{equation}
S_{r,t}[u]=\frac{1}{Q}\sum_{q=1}^{Q}|d_{q,r,t}[u]|^2,
\qquad Q=4.
\label{eq:power_spectrum}
\end{equation}
Thus, $d_{q,r,t}[u]$ is one complex DFT coefficient, whereas $S_{r,t}[u]$ is the measured power ordinate supplied to the inversion. The four channels observe the same particle ensemble. Their powers are therefore averaged, but the channels are not counted as four independent likelihood realizations.

{The range--Doppler illustrations use a recursive estimate of the coherent background to display the moving return (the non-stationary target returns) over the complete range interval. This operation is used only to form these illustrations:}
\begin{align}
y_{q,r,t}^{(\mathrm{vis})}[n]
&=y_{q,r,t}[n]-c_{q,r,t-1}[n],\\
c_{q,r,t}[n]
&=\alpha c_{q,r,t-1}[n]+(1-\alpha)y_{q,r,t}[n],
\qquad \alpha=0.9.
\label{eq:visual_background}
\end{align}
Here, $c_{q,r,t}[n]$ is a recursive estimate of the coherent part of the range-resolved signal $y_{q,r,t}[n]$. {With the \SI{50}{ms} frame interval, $\alpha=0.9$ gives this average an exponentially decaying memory with a relaxation time of approximately \SI{0.48}{s}.} A slow-time Hann window and a 256-point DFT then provide a smooth display. Neither the recursive estimate, the slow-time Hann window, nor the zero padding enters the Whittle quasi-likelihood, the Doppler moments, or the information criteria.

\subsection{Range-dependent background and finite-record Doppler spectrum}
The background term $b_{q,r,t}[n]$ in \eqref{eq:field_sum} is obtained from the no-motion measurement. The interval from 6 to \SI{27}{s} contains $N_b=421$ recorded acquisitions, whose average periodogram defines the reference background shape,
\begin{equation}
B_r[u]=\frac{1}{N_b}\sum_{t\in\mathcal T_b}S_{r,t}[u],
\qquad |\mathcal T_b|=N_b,
\label{eq:background_template}
\end{equation}
where $\mathcal T_b$ is the set of background acquisitions. The spectrum $B_r[u]$ is the frequency-domain counterpart of the background covariance $\mathbf C_{b,r}$ in \eqref{eq:data_model}. It retains the central stationary Doppler lobe and the {frequency-dependent, range-dependent structure} measured before fluidization. {Frequency dependent means that the background power varies across Doppler frequency rather than remaining spectrally flat.} Every candidate spectrum contains $a_{r,t}B_r[u]$, so a change in background strength is represented by $a_{r,t}$ rather than forced into the extended Doppler-lobe parameters.

A finite slow-time record supplies only one realization of the random return. Its periodogram is therefore jagged and changes between frames even when the underlying motion statistics are similar. An isolated peak is not automatically a separate particle stream. The estimator consequently fits the expected finite-record periodogram of the background-plus-particle model.

The first Doppler moments describe each range cell with one total power, one mean radial velocity, and one spectral width. This compact description is sufficient when one dominant extended Doppler lobe represents the local spectrum. In a bubbling bed, however, the same depth cell can contain appreciable contributions from particle ensembles moving towards and away from the radar. Their combined return may form more than one distinguishable extended Doppler lobe even when the overall mean remains near zero. The integer $K$ therefore counts the extended Doppler lobes used to represent one local spectrum and is subsequently called the spectral model order: $K=0$ denotes the scaled background alone, $K=1$ gives one effective ensemble-velocity distribution, and $K=2$ or 3 permits additional extended Doppler lobes. A value $K=2$ is interpreted as two distinguishable spectral contributions, not automatically as two independently resolved particle populations.

Using the centered normalized Doppler frequency $\widetilde f_u$ defined in \eqref{eq:centered_frequency_axis}, with $v_u=\lambda\widetilde f_u/(2T_s)$, the parameter set for a candidate containing $K\geq1$ extended Doppler lobes is
\begin{equation}
\boldsymbol{\Theta}_{K,r,t}
=\left\{
\widetilde\mu_{\kappa,r,t},\widetilde\sigma_{\kappa,r,t},P_{\kappa,r,t}
\right\}_{\kappa=1}^{K},
\label{eq:mixture_parameters}
\end{equation}
where $\kappa=1,\ldots,K$ is the lobe index, and $\widetilde\mu_{\kappa,r,t}$, $\widetilde\sigma_{\kappa,r,t}$, and $P_{\kappa,r,t}$ are the normalized mean Doppler frequency, normalized Doppler width, and power of lobe $\kappa$, respectively. {Following the finite-record parametric spectrum estimator (PSE) in \cite{Dash2024DopplerModel}, define the expected finite-record shape of one extended Doppler lobe as}

\begin{align}
G_N(\widetilde f_u;\widetilde\mu,\widetilde\sigma)
&=1+2\sum_{h=1}^{N-1}
\left(1-\frac{h}{N}\right)\nonumber\\
&\quad\times
\exp\!\left(-2\pi^2\widetilde\sigma^2h^2\right)
\cos\!\left(2\pi h(\widetilde\mu-\widetilde f_u)\right).
\label{eq:finite_lobe_kernel}
\end{align}
The background-only and $K$-lobe expected periodograms are then
\begin{align}
F_0(\widetilde f_u;a_{r,t})
&=a_{r,t}B_r[u],
\label{eq:background_model}\\
F_K(\widetilde f_u;\boldsymbol{\Phi}_{K,r,t})
&=a_{r,t}B_r[u]\nonumber\\
&\quad+\sum_{\kappa=1}^{K}P_{\kappa,r,t}
G_N\!\left(
\widetilde f_u;\widetilde\mu_{\kappa,r,t},\widetilde\sigma_{\kappa,r,t}
\right).
\label{eq:finite_mixture_spectrum}
\end{align}

Here, $S_{r,t}[u]$ in \eqref{eq:power_spectrum} is the measured periodogram ordinate, whereas $F_K(\widetilde f_u;\boldsymbol{\Phi}_{K,r,t})$ is its modelled expectation under candidate order $K$. The parameter vector $\boldsymbol{\Phi}_{K,r,t}=\{a_{r,t},\boldsymbol{\Theta}_{K,r,t}\}$ contains the background scale and all lobe parameters. In \eqref{eq:finite_lobe_kernel}, $h$ is sweep lag and $N=128$ is the finite record length. The factor $1-h/N$ accounts for the finite rectangular observation interval. Equations~\eqref{eq:finite_lobe_kernel} and \eqref{eq:finite_mixture_spectrum} therefore describe the expected finite-record periodogram rather than an ideal continuous Gaussian curve.

All 128 non-oversampled ordinates of the DFT are fitted:
\begin{equation}
\mathcal U_f=
\left\{0,\ldots,N-1\right\},
\qquad N_{\mathrm{fit}}=|\mathcal U_f|=128.
\label{eq:fit_bins}
\end{equation}
The complete unambiguous interval constrains both the measured {frequency-dependent background} and the fitted extended Doppler-lobe spectrum. This is important near zero velocity: the central stationary contribution is represented by the scaled background $a_{r,t}B_r[u]$, while any additional broadening or displacement that cannot be explained by that background is assigned to the extended Doppler-lobe term. Zero-padded display samples are not counted as independent observations.

\subsection{Doppler-spectrum detectability}
The one-lobe retrieval is first performed in every evaluated range--time cell. This produces a candidate extended Doppler lobe even when the measured spectrum is adequately represented by the background alone. A separate decision is therefore made before the fitted lobe is interpreted as particle motion. Every cell is represented by the scaled background plus one extended Doppler lobe, i.e., the $K=1$ model. Let
\begin{equation}
\widehat S^{(\mathrm{lobe})}_{1,r,t}[u]
=F_1(\widetilde f_u;\widehat{\boldsymbol{\Phi}}_{1,r,t})
-\widehat a_{r,t}B_r[u]
\label{eq:particle_spectrum}
\end{equation}
denote the fitted extended Doppler-lobe contribution, where $\widehat{\boldsymbol{\Phi}}_{1,r,t}$ is the one-lobe estimate and $\widehat a_{r,t}$ is its fitted background-scale entry. Its power relative to the fitted background {in decibels} is written as
\begin{equation}
\Gamma_{r,t}
=10\log_{10}
\frac{\displaystyle\sum_{u\in\mathcal U_f}\widehat S^{(\mathrm{lobe})}_{1,r,t}[u]}
{\displaystyle\sum_{u\in\mathcal U_f}\widehat a_{r,t}B_r[u]}.
\label{eq:particle_contrast}
\end{equation}
Thus, $\Gamma_{r,t}=0$ dB means that the candidate lobe and the scaled background have equal integrated power; a negative value means that the candidate lobe is weaker. The comparison is made separately at every range because the stationary return is not equally strong throughout the vessel.

The no-motion interval shows how large $\Gamma_{r,t}$ can become when no fluidization-induced Doppler spectrum is present. For each range $r$, the detection threshold $\tau_r$ is its 99th percentile, estimated from every fifth no-motion acquisition and pooled with the two neighboring ranges on either side. An interior range therefore uses approximately 425 reference values. A range--time cell is declared Doppler-detectable when
\begin{equation}
d_{r,t}=
\begin{cases}
1, & \Gamma_{r,t}>\tau_r,\\
0, & \Gamma_{r,t}\leq\tau_r,
\end{cases}
\label{eq:motion_gate}
\end{equation}
where $d_{r,t}=1$ denotes a detected cell and $d_{r,t}=0$ a cell that is not detected. Let $\mathcal R_0$ contain the 102 evaluated apparent-range cells from \SI{2.518}{m} to \SI{4.489}{m}; the detected ranges at time $t$ are $\mathcal R_t=\{r\in\mathcal R_0:d_{r,t}=1\}$. Accordingly, a cell is retained only when its fitted extended-lobe power exceeds what the no-motion measurements normally produce at that range. The 99th-percentile choice allows approximately one in one hundred no-motion reference values to cross the threshold by construction; the measured exceedance rate is 0.77\% within the calibration interval. 

The threshold sets a direct trade-off. Increasing it produces cleaner maps and rejects more stationary-background fluctuations, but weak particle motion can then be missed. Decreasing it makes the radar more sensitive to weak motion, but more background variations are retained. A filtered range cell in a filtered map, therefore, means that a particle-motion spectrum cannot be distinguished from the calibrated background during that short record; it does not mean that the physical cell contains no particles. Widths below the independent Doppler-bin spacing are retained and marked as unresolved. Only cells in $\mathcal R_t$ enter the subsequent comparison of one, two, or three extended Doppler lobes; the initial moment maps are not screened by an information criterion.
\subsection{Cost function development; Whittle log-likelihood}
The covariance model in \eqref{eq:data_model} describes how both the particle return and measured background retain memory over sweep lag. For candidate order $K$, let $\mathbf C_{s,K,r,t}$ be the sum of the $K$ lobe covariance matrices and define the corresponding total covariance as $\mathbf C_{K,r,t}=\mathbf C_{s,K,r,t}+a_{r,t}\mathbf C_{b,r}$. Direct evaluation of the proper-complex Gaussian likelihood requires the determinant and inverse of this matrix in every range--time cell. For one receive channel, its covariance-form log-likelihood is
\begin{equation}
\begin{aligned}
\mathcal L_{\mathrm C,K,q,r,t}
={}&-\log\det\mathbf C_{K,r,t}\\
&-\mathbf y_{q,r,t}^{H}
\mathbf C_{K,r,t}^{-1}
\mathbf y_{q,r,t}
+\mathrm{constant}.
\end{aligned}
\label{eq:complex_gaussian_likelihood}
\end{equation}
{Here, the superscript $H$ denotes the Hermitian transpose of a vector.}
The same second-order model can be expressed in frequency through the Doppler spectrum. Define the centered DFT vector by stacking the coefficients in \eqref{eq:doppler_fft},
\begin{equation}
\begin{aligned}
\mathbf d_{q,r,t}
&=\begin{bmatrix}
d_{q,r,t}[0]&\cdots&d_{q,r,t}[N-1]
\end{bmatrix}^{T},\\
\mathbf d_{q,r,t}
&=\mathbf G\mathbf y_{q,r,t},\\
\boldsymbol{\Sigma}_{d,K,r,t}
&=\mathbf G\mathbf C_{K,r,t}\mathbf G^{H}.
\end{aligned}
\label{eq:frequency_covariance}
\end{equation}
where $\mathbf G$ is the centered rectangular-record $N$-point DFT matrix. Thus, lowercase $d_{q,r,t}[u]$ is one frequency coefficient, bold $\mathbf d_{q,r,t}$ collects all coefficients, and $\boldsymbol{\Sigma}_{d,K,r,t}$ is their covariance under candidate order $K$.
Using the vectorization identity,
\begin{equation}
\operatorname{vec}(\boldsymbol{\Sigma}_{d,K,r,t})
=\left(\mathbf G^{*}\otimes\mathbf G\right)
\operatorname{vec}(\mathbf C_{K,r,t}),
\label{eq:kronecker_frequency_covariance}
\end{equation}
where $\operatorname{vec}(\cdot)$ stacks the columns of a matrix and $\otimes$ is the Kronecker product. Equation~\eqref{eq:kronecker_frequency_covariance} makes explicit that the frequency-domain covariance is the Fourier-coordinate representation of the same time-domain covariance.

For a stationary process, the DFT approximately diagonalizes the Toeplitz covariance matrix as the record length increases. In physical terms, separated Doppler ordinates are then treated as approximately uncorrelated descriptions of how the random signal power is distributed in frequency. The Whittle approximation retains the diagonal entries,
\begin{equation}
\boldsymbol{\Sigma}_{d,K,r,t}
\approx
\operatorname{diag}_{u\in\mathcal U_f}\!\left\{
F_K(\widetilde f_u;\boldsymbol{\Phi}_{K,r,t})
\right\}.
\label{eq:whittle_diagonal}
\end{equation}
When the off-diagonal terms are negligible, the DFT coefficients are uncorrelated; because they are proper complex Gaussian, they are then independent. Their squared magnitudes are exponentially distributed about the expected periodogram \cite{Levin1965PowerEstimation}. Substituting this diagonal approximation into the Gaussian likelihood gives the Whittle likelihood \cite{Whittle1953,Frehlich1993,Levin1965PowerEstimation,Dash2024DopplerModel}. The finite-record expression in \eqref{eq:finite_mixture_spectrum} retains the expected leakage and broadening, whereas the approximation neglects the remaining correlation between neighbouring periodogram bins. The four receivers view the same particle ensemble, so their powers are averaged to {smooth} the measured spectrum but are not counted as independent realizations. Power averaging changes the exact exponential distribution. The objective is consequently used as a Whittle quasi-likelihood: it preserves the physically derived expected spectrum while acknowledging that the assumed probability law is approximate.

The negative Whittle log-likelihood for candidate $K$ is
\begin{equation}
\begin{aligned}
\mathcal J_K(\boldsymbol{\Phi}_{K,r,t})
=\sum_{u\in\mathcal U_f}
\Bigg[
&\log\!\left(
\pi F_K(\widetilde f_u;\boldsymbol{\Phi}_{K,r,t})
\right)\\
&+\frac{S_{r,t}[u]}
{F_K(\widetilde f_u;\boldsymbol{\Phi}_{K,r,t})}
\Bigg].
\end{aligned}
\label{eq:whittle_cost}
\end{equation}
For $K=0$, the same expression is minimized over the background scale $a_{r,t}$. For $K\geq1$, the scale and extended Doppler-lobe parameters are estimated jointly:
\begin{equation}
\begin{aligned}
\widehat{\boldsymbol{\Phi}}_{K,r,t}
&=\arg\min_{\boldsymbol{\Phi}_{K,r,t}}
\mathcal J_K(\boldsymbol{\Phi}_{K,r,t}),\\
\widehat{\mathcal L}_{K,r,t}
&=-\mathcal J_K(
\widehat{\boldsymbol{\Phi}}_{K,r,t}).
\end{aligned}
\label{eq:whittle_estimator}
\end{equation}
The $K=2$ optimization starts both from the fitted $K=1$ spectrum augmented by the strongest remaining spectral lobe and from a separated-lobe initialization; $K=3$ is initialized analogously from $K=2$. 

During optimization, each normalized width is re-parametrized as $\widetilde\sigma_{\kappa}=\exp(\xi_\kappa)$ and the unrestricted real variable $\xi_\kappa$ is optimized. A signed width cannot be identified because the covariance and spectrum depend on $\widetilde\sigma_{\kappa}^{2}$; allowing positive and negative widths would duplicate the same solution. The logarithmic parameter lets the optimizer approach arbitrarily narrow positive values without imposing a lower bound at one Doppler bin. The higher-order solution is retained only when its maximized likelihood is not smaller than that of the nested lower-order model. The fitted lobes are ordered by mean velocity. Lobe 1 therefore denotes the lowest-velocity contribution and is not a permanently assigned physical particle population. The numerical minimization uses the active-set and Limited Memory Broyden--Fletcher--Goldfarb--Shanno (L-BFGS) algorithms \cite{Liu2007ConvergenceProblems,Broyden1970TheConsiderations,MatlabOTB}.

\subsection{Estimating the number of Doppler lobes}
The Akaike information criterion (AIC) and Bayesian information criterion (BIC) compare the improvement in spectral fit with the number of estimated lobe parameters \cite{Akaike1974,Schwarz1978}. They are applied only after the extended Doppler-lobe spectrum is detectable and compare $K\in\{1,2,3\}$; they do not decide whether a range cell contains measurable motion. For each candidate,
\begin{align}
\mathrm{AIC}_{K,r,t}
&=2p_K-2\widehat{\mathcal L}_{K,r,t},\\
\mathrm{BIC}_{K,r,t}
&=p_K\log N_{\mathrm{fit}}
-2\widehat{\mathcal L}_{K,r,t},
\label{eq:aic_bic}
\end{align}
where $p_K=1+3K$ because every candidate estimates one background scale in addition to three parameters per extended Doppler lobe. The fitted-sample count is $N_{\mathrm{fit}}=128$ from \eqref{eq:fit_bins}. The Whittle approximation treats these non-oversampled Fourier coefficients as approximately independent. The AIC and BIC are used as relative diagnostics among the candidate expected spectra. Their selected fractions are not posterior probabilities and do not by themselves prove the existence of separate physical particle populations. However, they estimate the number of distinguishable extended Doppler lobes in the measured spectrum and therefore indicate whether more than one ensemble velocity centre is supported. 

For each cell $r\in\mathcal R_t$, the BIC-selected number of extended Doppler lobes is
\begin{equation}
\widehat K_{r,t}^{(\mathrm{BIC})}
=\arg\min_{K\in\{1,2,3\}}
\mathrm{BIC}_{K,r,t}.
\label{eq:bic_order}
\end{equation}
AIC uses the same background-calibrated detectability set $\mathcal R_t$ and provides a less strongly penalized comparison among $K=1$, 2, and 3. The log-likelihood and BIC gains from one to two extended Doppler lobes are
\begin{align}
\Delta\mathcal L_{21,r,t}
&=\widehat{\mathcal L}_{2,r,t}
-\widehat{\mathcal L}_{1,r,t},\\
\Delta\mathrm{BIC}_{21,r,t}
&=\mathrm{BIC}_{1,r,t}
-\mathrm{BIC}_{2,r,t}\nonumber\\
&=2\Delta\mathcal L_{21,r,t}
-3\log N_{\mathrm{fit}}.
\label{eq:bic_gain}
\end{align}
A positive $\Delta\mathrm{BIC}_{21,r,t}$ means that the improvement supplied by the second extended Doppler lobe exceeds the BIC penalty for its three additional parameters. The same definitions apply to the gain from $K=2$ to $K=3$.

\subsection{Radar quantities for pressure comparison and directional motion}
At time $t$, the set $\mathcal R_t$ contains the range cells in which the fitted Doppler return is distinguishable from the measured no-motion background. The one-lobe fit provides three radar quantities in every cell $r\in\mathcal R_t$: fitted total power $\widehat P(r,t)$, mean radial velocity $\widehat\mu_v(r,t)$, and Doppler spectral width $\widehat\sigma_v(r,t)$. The hat denotes a quantity estimated from the radar measurement. Because all three quantities in this subsection come from the $K=1$ fit, no additional lobe index is required.

Pressure describes the bulk state of the bed, whereas the radar quantities vary with range. Two summaries are used to compare them. First, the fitted power is summed over the detected bed layer,
\begin{equation}
P_{\mathrm{bed}}(t)
=\sum_{r\in\mathcal R_t}\widehat P(r,t).
\label{eq:k1_range_power}
\end{equation}
This quantity increases when the detectable moving layer occupies more range cells, when the return becomes stronger, or both. Second, the representative local Doppler width is
\begin{equation}
\overline{\sigma}_v(t)
=\frac{\displaystyle\sum_{r\in\mathcal R_t}
\widehat P(r,t)\widehat\sigma_v(r,t)}
{\displaystyle P_{\mathrm{bed}}(t)}.
\label{eq:k1_mean_width}
\end{equation}
The weighting gives greater influence to range cells with stronger particle return. Consequently, $\overline{\sigma}_v(t)$ is the mean of the local fitted widths; it is not the width of one spectrum formed by combining the complete bed.

The pressure record is interpolated to the radar time instants, which are separated by \SI{0.05}{s}. An instantaneous radar summary is formed when at least five detectable range cells are available. The displayed curves use a 5-s running median and remain blank when fewer than half of the radar records in that interval satisfy this requirement.

Let $p(t)$ denote the displayed pressure sequence, and let $g(t)$ denote one radar summary at a time: either $10\log_{10}P_{\mathrm{bed}}(t)$ or $\overline{\sigma}_v(t)$. The symbol $g$ is used here to avoid reusing $z$, which already denotes the range-transform output in \eqref{eq:range_fft}. Their linear association is quantified by the Pearson correlation coefficient
\begin{equation}
\rho_{p,g}=
\frac{\displaystyle\sum_{t\in\mathcal T}
\big(p(t)-\overline p\big)
\big(g(t)-\overline g\big)}
{\displaystyle
\sqrt{\sum_{t\in\mathcal T}\big(p(t)-\overline p\big)^2}
\sqrt{\sum_{t\in\mathcal T}\big(g(t)-\overline g\big)^2}},
\label{eq:pearson}
\end{equation}
where $\mathcal T$ contains the times from 0.10 to \SI{8}{min} for which both 5-s median sequences are available, and $\overline p$ and $\overline g$ are their averages over $\mathcal T$. A positive coefficient means that pressure and the chosen radar quantity generally increase together over the imposed operating sequence; a negative coefficient means that one generally decreases as the other increases. The coefficient describes association and does not imply an instantaneous pressure-to-velocity relation.

The sign of the fitted mean velocity separates motion towards and away from the radar. Define
\begin{align}
\mathcal R_-(t)
&=\left\{r\in\mathcal R_t:
\widehat\mu_v(r,t)<0\right\},\\
\mathcal R_+(t)
&=\left\{r\in\mathcal R_t:
\widehat\mu_v(r,t)>0\right\}.
\label{eq:k1_directional_sets}
\end{align}
Negative mean velocity denotes motion towards the top-mounted radar, and positive mean velocity denotes motion away from it. For either sign, the representative velocity is the power-weighted mean
\begin{equation}
\overline\mu_{\pm}(t)
=\frac{\displaystyle\sum_{r\in\mathcal R_{\pm}(t)}
\widehat P(r,t)\widehat\mu_v(r,t)}
{\displaystyle\sum_{r\in\mathcal R_{\pm}(t)}
\widehat P(r,t)}.
\label{eq:k1_directional_means}
\end{equation}
The fraction of the detected Doppler-lobe power carried by that direction is
\begin{equation}
\eta_{\pm}(t)
=\frac{\displaystyle\sum_{r\in\mathcal R_{\pm}(t)}
\widehat P(r,t)}
{\displaystyle P_{\mathrm{bed}}(t)}.
\label{eq:k1_directional_fractions}
\end{equation}
The directional quantities are evaluated only when the corresponding range set is nonempty. They show how the detected power is divided between positive- and negative-mean range cells; they do not assume that two extended Doppler lobes coexist within one cell. The range--time map retains the recorded sampling, while the displayed directional curves use a \SI{0.5}{s} running median.
\section{Results}\label{sec:results}
\subsection{Range-Doppler observations}
\begin{figure}[t]
\centering
\includegraphics[width=0.98\textwidth]{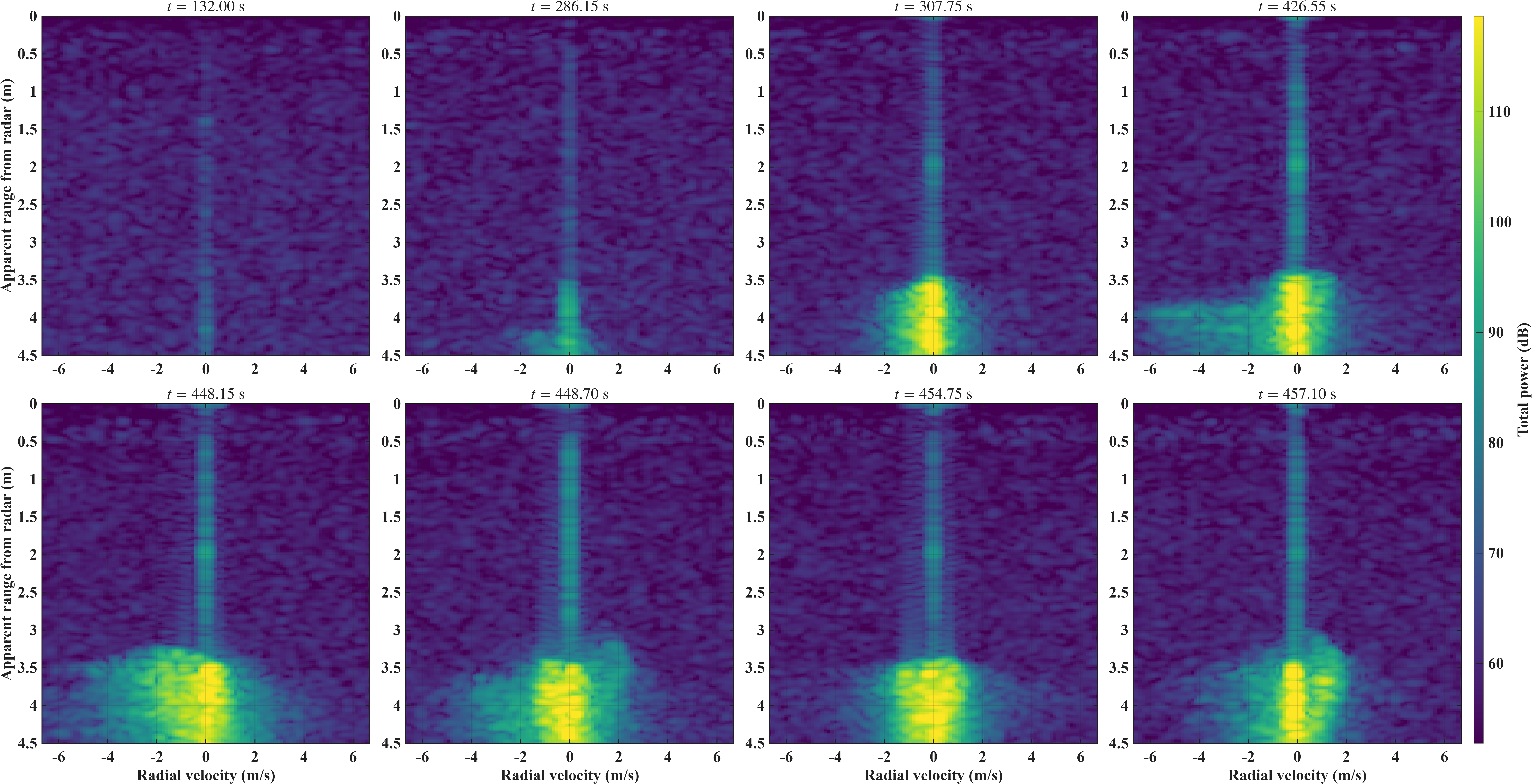}
\caption{Range-Doppler total power at $t=132$, 286.15, 307.75, 426.55, 448.15, 448.70, 454.75, and 457.10 s. The spectra use fast-time Hann windowing, recursive coherent-background removal, slow-time Hann windowing, and a 256-point Doppler transform. All panels use one common dB scale and the viridis colour map; no panel-wise power normalization is applied. Apparent range is displayed from the radar to \SI{4.5}{m}; delayed-path responses beyond this physical analysis limit are omitted. The velocity interval, \SIrange{-6.68}{6.68}{m/s}, is a zoom within the full unambiguous interval of approximately $\pm\SI{17.096}{m/s}$.}
\label{fig:rd}
\end{figure}

Fig.~\ref{fig:rd} shows how the range-resolved velocity distribution changes at eight times spanning the imposed operating sequence. Apparent range increases downwards from the top-mounted radar, so energy extending towards smaller range indicates motion and bed expansion towards the radar. The display stops at \SI{4.5}{m}; delayed returns at larger apparent range are excluded from the physical interpretation. The figure uses the recursive background suppression defined in \eqref{eq:visual_background}, whereas the statistical retrieval uses the measured background shape in \eqref{eq:background_template} with the cell-dependent scale in \eqref{eq:background_model}.

The particle return evolves from weak echoes near the lower vessel to strong Doppler power extending towards smaller range. Several panels contain power on both sides of zero radial velocity. The measurements therefore indicate that particle ensembles move both towards and away from the radar, although the number of statistically distinguishable contributions cannot be assigned by visual inspection alone.

\subsection{Measured no-motion background}
\begin{figure}[t]
\centering
\includegraphics[width=0.98\textwidth]{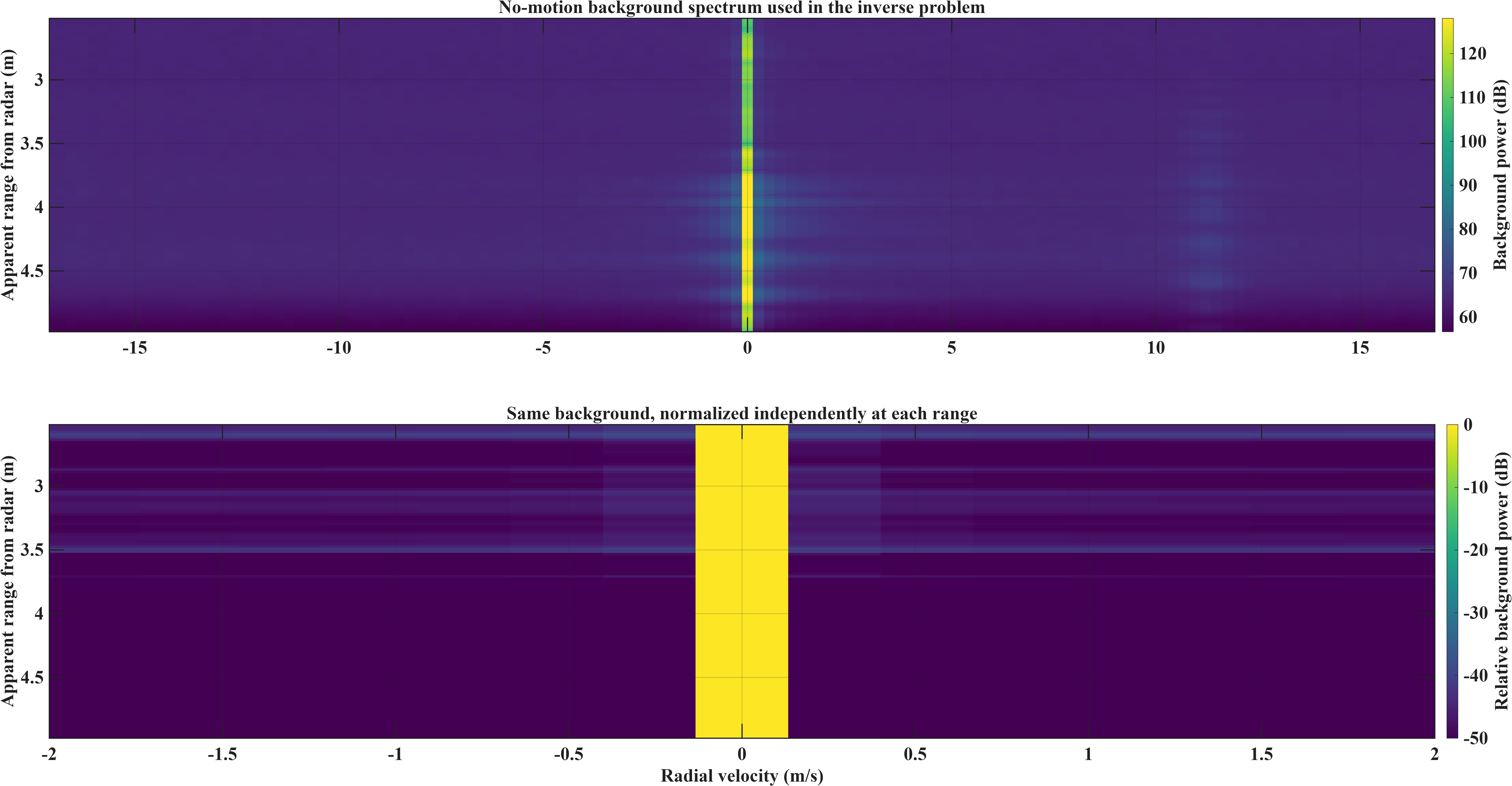}
\caption{Measured no-motion range--Doppler background $B_r[u]$, averaged from 6 to \SI{27}{s} and used in the spectral inversion. The upper panel shows the absolute background power spectral density over the full unambiguous velocity interval. The lower panel shows the same data over \SIrange{-2}{2}{m/s}, after each range profile has been divided by its own maximum for display only. This normalization exposes the narrow stationary response near zero velocity and the range-dependent spectral shape; it is not used in the inversion. Both panels use the viridis colour map, and range increases downwards from the radar.}
\label{fig:background_rd}
\end{figure}

The spectral background used by the inversion is shown in Fig.~\ref{fig:background_rd}. It is the power spectral density (PSD) averaged over the no-motion interval in \eqref{eq:background_template}; it is not a fitted analytical clutter curve and is not subtracted from the data. The upper panel retains its absolute power and shows that the stationary return varies strongly with range. The lower panel divides each range profile by its own maximum only to make the spectral shape visible. It reveals a narrow response around zero radial velocity together with fixed range-dependent structure. The inversion uses the absolute background in the upper panel and estimates only its scale $a_{r,t}$ for each later range--time cell.

This background measurement is essential for interpretation. A peak that is already present before fluidization cannot by itself be assigned to moving particles. Particle motion is detected only when the fitted extended Doppler lobe rises above the range-dependent level established by these no-motion spectra.

\subsection{One-lobe Doppler retrieval}
\begin{figure}[t]
\centering
\includegraphics[width=0.98\textwidth]{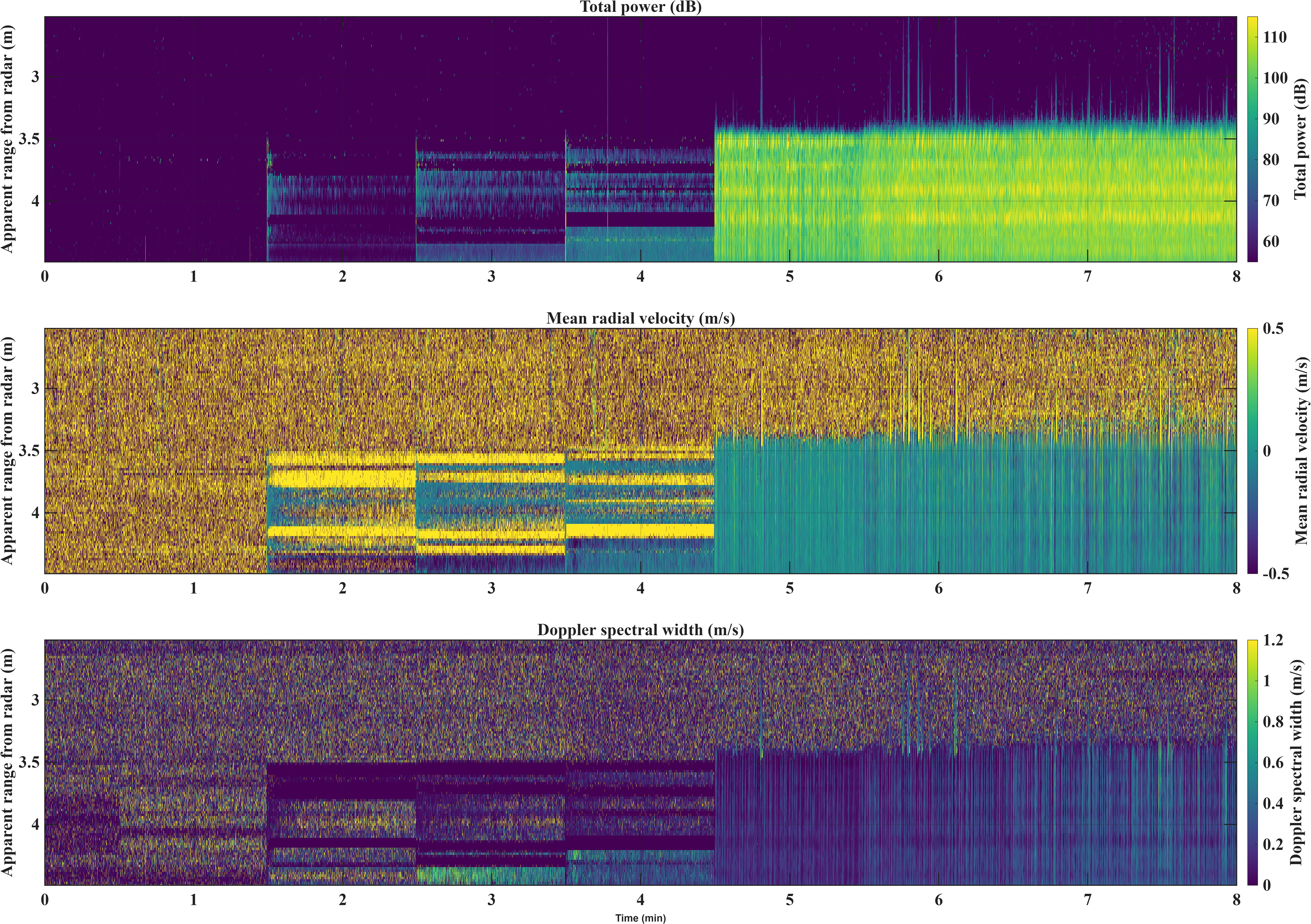}
\caption{Complete-record $K=1$ Whittle estimates before Doppler-spectrum detectability. One extended Doppler lobe is fitted wherever the numerical optimization converges, including cells dominated by the measured background. The total-power limits are identical to those in Fig.~\ref{fig:k1_full_maps_multi}; mean velocities outside \SIrange{-0.2}{0.2}{m/s} and widths outside \SIrange{0}{0.7}{m/s} are clipped at the displayed colour limits. These unscreened maps show the inversion output but are not, by themselves, particle-motion measurements.}
\label{fig:k1_full_maps_unthresholded}
\end{figure}

Figure~\ref{fig:k1_full_maps_unthresholded} retains the $K=1$ result in every range--time cell before the detectability decision. It therefore shows both the response of the fluidized bed and the values returned when the measured spectrum is dominated by the background in Fig.~\ref{fig:background_rd}. The fitted total power, mean radial velocity, and spectral width are physical particle-motion quantities only after the extended Doppler lobe is strong enough to be detected.

The horizontal bands in the unscreened mean-velocity and width maps illustrate why the power decision is required. In the prominent band near \SIrange{4.10}{4.20}{m} between 3.5 and \SI{4.5}{min}, representative fits assign more than \SI{160}{dB} less power to the candidate extended Doppler lobe than to the scaled background. Because that lobe contributes virtually no measured power, its fitted centre and width are not meaningful particle-motion estimates; clipping them to the displayed colour interval makes the band appear visually strong. Equation~\eqref{eq:motion_gate} rejects these cells.

\begin{figure}[t]
\centering
\includegraphics[width=0.98\textwidth]{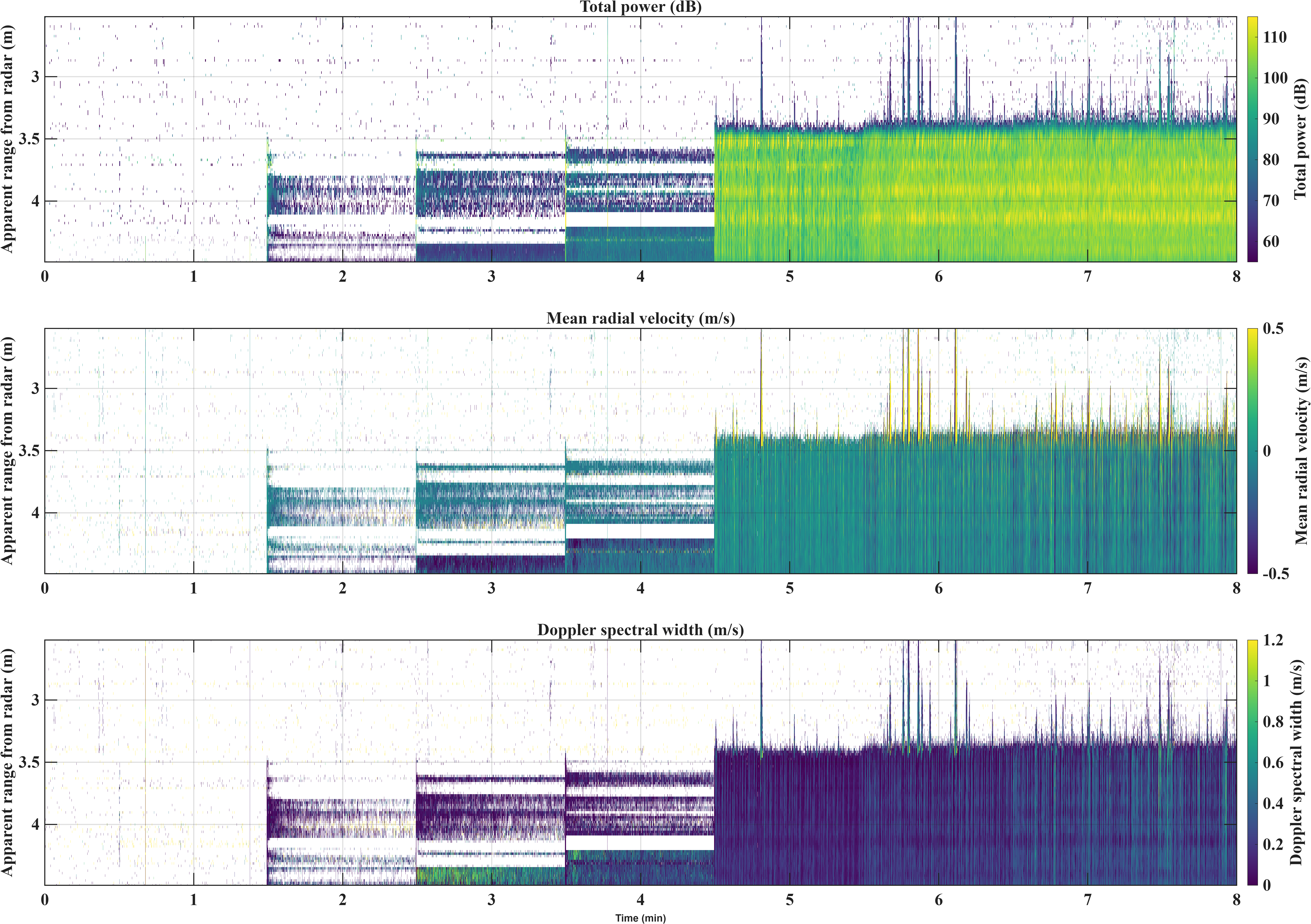}
\caption{Complete-record $K=1$ Whittle retrievals after background-calibrated Doppler-spectrum detectability: fitted total power, mean radial velocity, and Doppler spectral width. Parameters are obtained at the recorded \SI{0.05}{s} interval. White cells do not exceed the calibrated extended-lobe-to-background contrast and are not evidence of an empty range cell. Range increases downwards from the radar. Mean velocities outside the displayed \SIrange{-0.2}{0.2}{m/s} interval are clipped at the colour limits; the displayed width interval is \SIrange{0}{0.7}{m/s}.}
\label{fig:k1_full_maps_multi}
\end{figure}

The background-calibrated detectability decision is then applied before testing additional extended Doppler lobes. Every range-time cell has first been represented by the $K=1$ finite-record spectrum containing a scaled background and one extended Doppler lobe. Fig.~\ref{fig:k1_full_maps_multi} shows the fitted total power, mean radial velocity, and Doppler spectral width where the extended Doppler-lobe contribution exceeds the range-dependent detectability threshold in \eqref{eq:motion_gate}. The detected Doppler spectrum first appears near the lower bed and later extends towards smaller range as the bed expands and particle motion becomes detectable higher in the vessel. Total power increases markedly during developed fluidization. The mean alternates around zero, whereas the width describes the velocity spread assigned to the local scattering ensemble by the one-lobe representation. White cells do not pass the calibrated Doppler-lobe-to-background contrast. They are not evidence of an empty range cell: particles can remain present when their return is stationary, is represented by the measured central background lobe, or cannot be separated from the background over the finite observation.

A weaker horizontal band can remain after this decision when a repeatable spectral change exceeds the no-motion threshold at one fixed range. Detectability then states only that the measured spectrum differs from its no-motion reference; it does not show that a constant-velocity particle layer exists. The physical interpretation therefore relies on structures that change over neighboring ranges and times, on their fitted power, and on the range--Doppler measurements. The bidirectional-motion result below is based on such evolving positive- and negative-mean regions rather than on an isolated horizontal band.

\subsection{Pressure and integrated radar quantities}
\begin{figure}[!t]
\centering
\includegraphics[width=0.98\textwidth]{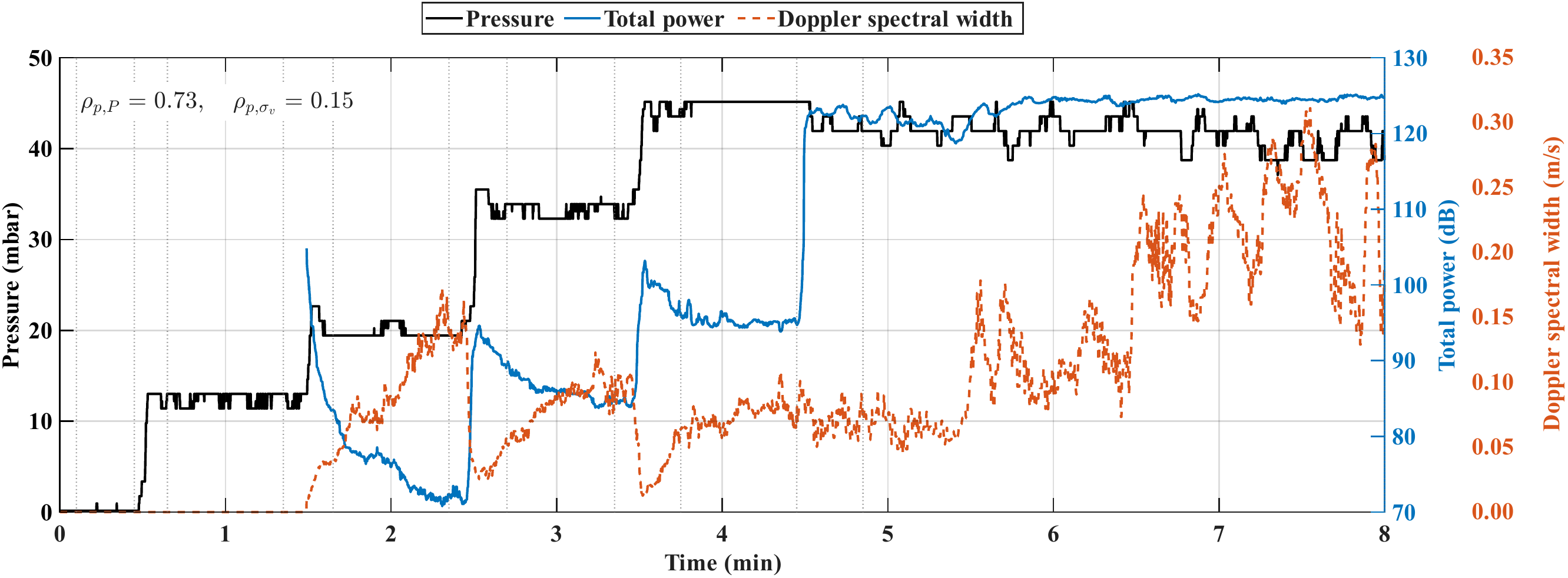}
\caption{Pressure, total Doppler-lobe power summed over the detected range cells, and power-weighted mean local Doppler spectral width over the complete sequence. The radar quantities use cells within \SIrange{2.518}{4.489}{m}. The three vertical axes retain their physical units; the displayed width axis is limited to \SIrange{0}{0.35}{m/s}, and larger values are clipped at the upper boundary. The curves show 5-s running medians. Each radar value requires at least five detected range cells, and a 5-s interval remains blank when fewer than half of its radar records satisfy this requirement. Vertical dotted lines delimit the stable pressure intervals. The Pearson coefficients use the finite, unclipped 5-s trends over 0.10--\SI{8}{min}.}
\label{fig:pressure_moments}
\end{figure}

Fig.~\ref{fig:pressure_moments} compares pressure with the radar total power and Doppler spectral width defined in \eqref{eq:k1_range_power} and \eqref{eq:k1_mean_width}. Pressure uses the left vertical axis, while the two radar quantities use separate right vertical axes. At the beginning of the sequence, particle motion is detectable in only a few range cells. As the bed becomes fluidized, the detected layer extends towards the radar and its total fitted power increases. Both radar curves are formed from the same detected range cells and the same one-lobe fits used throughout the results.

The 5-s median trends give $\rho_{p,P}=0.734$ and $\rho_{p,\sigma_v}=0.152$ over 0.10--\SI{8}{min}. The stronger power correlation means that the overall strength and vertical extent of the detected radar return generally increase with the pressure-defined fluidization state. The width correlation is weaker because Doppler width describes the spread of radial velocities inside each detected range cell. It is influenced by local particle motion and need not increase in direct proportion to the pressure of the complete bed. Pressure and radar power therefore follow the imposed operating sequence together, while Doppler width supplies different information about the local motion.

\begin{figure}[!t]
\centering
\includegraphics[width=0.98\textwidth]{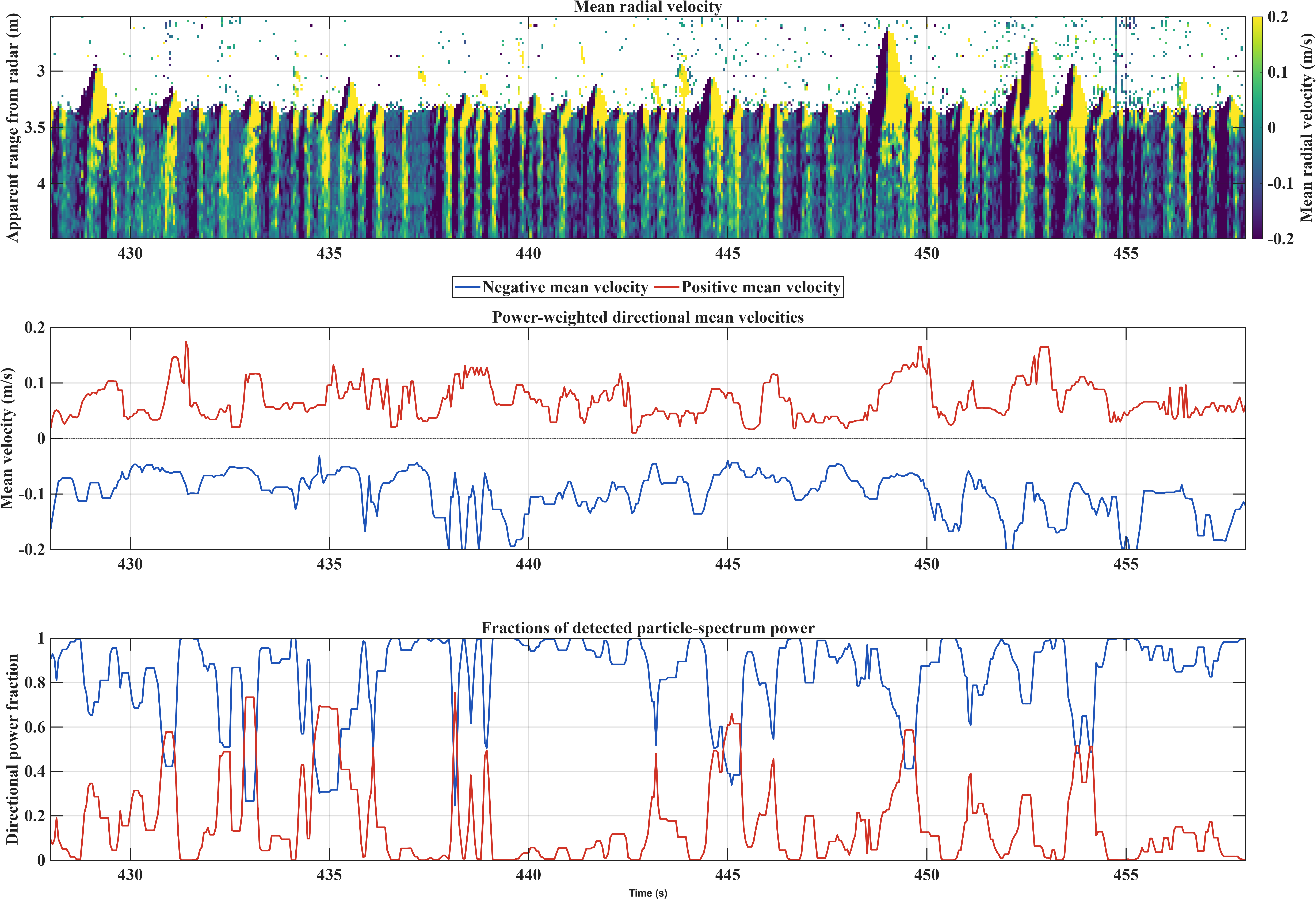}
\caption{Bidirectional $K=1$ Doppler retrieval from 428 to \SI{458}{s}. The upper panel shows the unsmoothed mean radial velocity over range and time; white cells do not exceed the calibrated extended-lobe-to-background contrast. The middle panel gives the power-weighted negative- and positive-mean velocities in \eqref{eq:k1_directional_means}, and the lower panel gives their power fractions in \eqref{eq:k1_directional_fractions}. The displayed curves use a \SI{0.5}{s} running median. Range increases downwards from the radar.}
\label{fig:bidirectional_motion}
\end{figure}

\subsection{Bidirectional particle motion}
The complete-record mean-Doppler map establishes that both velocity signs occur throughout the developed fluidization interval. Fig.~\ref{fig:bidirectional_motion} resolves this behaviour from 428 to \SI{458}{s}, while the pressure in Fig.~\ref{fig:pressure_moments} remains within the final operating condition. The unsmoothed range-time map contains connected positive- and negative-mean regions that repeatedly appear, disappear, and change sign across neighbouring ranges. The power-weighted curves in \eqref{eq:k1_directional_means} remain on opposite sides of zero, and the fractions in \eqref{eq:k1_directional_fractions} show how the detected extended-lobe power is redistributed between the two radial directions. These quantities use the fitted $K=1$ spectra and therefore establish bidirectional motion over range and time without presuming that two extended Doppler lobes coexist within an individual cell.

Positive and negative means at different ranges show that the dominant local motion occurs in both radial directions. Their irregular alternation and the nonzero spectral widths in Fig.~\ref{fig:k1_full_maps_multi} are consistent with bubbling-induced convective circulation and velocity fluctuations. The radar observations alone do not constitute a complete turbulence measurement because they contain only the radial component within a broad angular footprint. This observation does not require two fitted lobes in every cell: $K=1$ describes the dominant local velocity, whereas $K\geq2$ asks whether distinct contributions coexist within the same local spectrum.

\subsection{Selection of the number of Doppler lobes; $K$}
The number of extended Doppler lobes is assessed only after the $K=1$ fit exceeds the calibrated detectability threshold. Within the 428-\SI{458}{s} interval, a Doppler spectrum is detectable in 63.42\% of the evaluated range-time cells. Conditional on this set, AIC selects $K=1$, 2, and 3 in 85.00\%, 13.17\%, and 1.83\% of the cells, respectively; BIC gives 96.64\%, 3.15\%, and 0.22\%. The comparisons therefore answer whether one extended Doppler lobe is sufficient or whether the local periodogram supports additional velocity centres; they do not determine whether particles occupy the range cell. AIC applies the smaller complexity penalty and consequently admits weaker departures from the one-lobe representation, whereas BIC requires stronger evidence for each additional lobe.

\begin{figure}[t]
\centering
\includegraphics[width=0.98\textwidth]{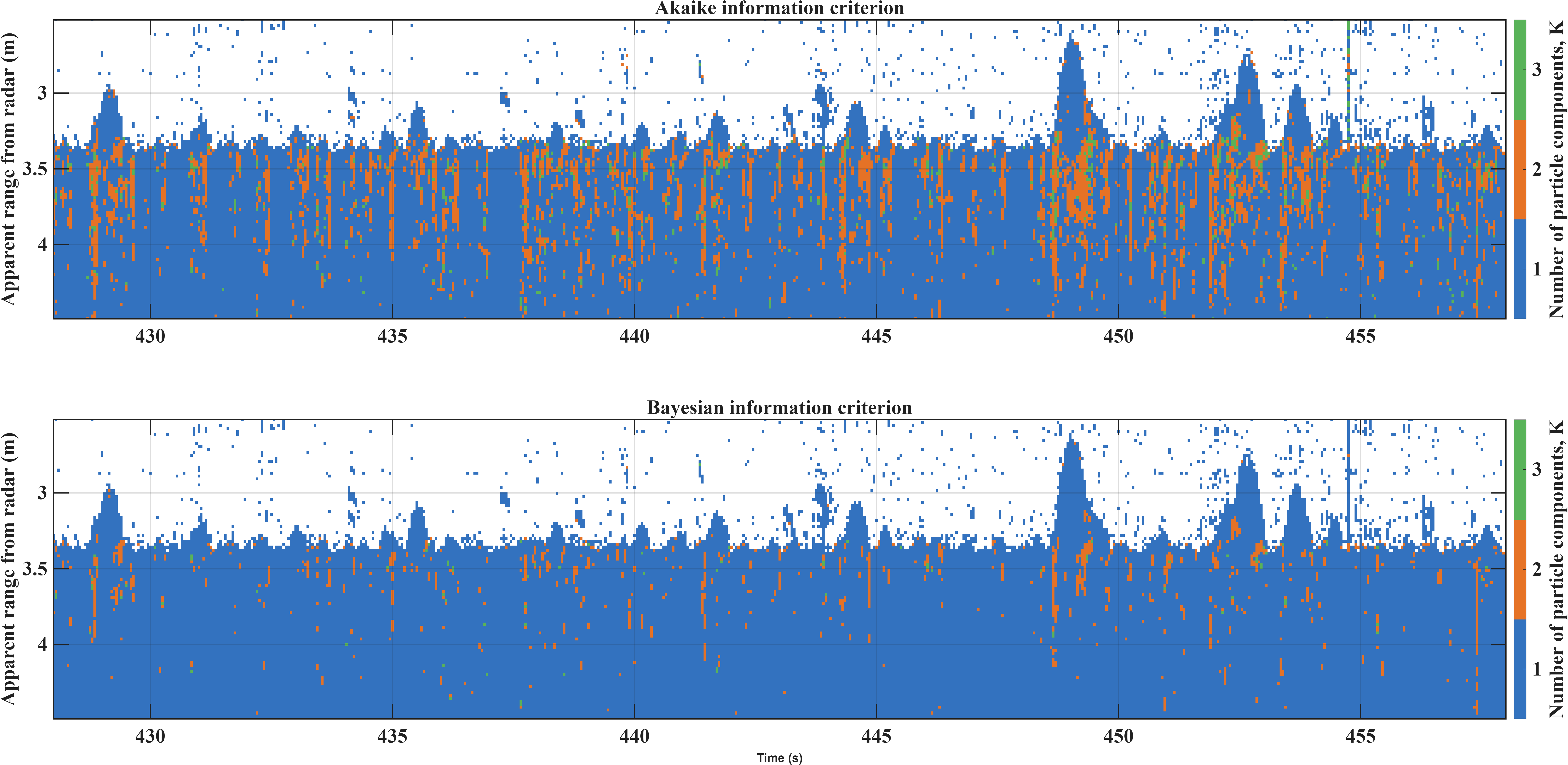}
\caption{Akaike-information-criterion (AIC)- and Bayesian-information-criterion (BIC)-selected Doppler-spectrum representations from 428 to \SI{458}{s}, evaluated only where the fitted $K=1$ extended Doppler lobe exceeds the calibrated lobe-to-background contrast within \SIrange{2.518}{4.489}{m}. White cells do not pass this detectability criterion; one, two, and three denote the selected numbers of extended Doppler lobes. Both panels use the scaled range-dependent background, the 128-point rectangular estimation spectrum, and all 128 fitted non-oversampled ordinates.}
\label{fig:aic_bic_order}
\end{figure}

For $N_{\mathrm{fit}}=128$, adding one extended Doppler lobe introduces three parameters and requires
\begin{equation}
\Delta\mathcal L_{K+1,K}>
\frac{3}{2}\log N_{\mathrm{fit}}
=7.28
\label{eq:bic_likelihood_threshold}
\end{equation}
to produce a positive BIC gain. Fig.~\ref{fig:aic_bic_order} shows the AIC and BIC decisions over the same interval used for the directional observation. Both criteria place additional lobes in localized range-time regions. BIC retains fewer such cells because the required likelihood improvement in \eqref{eq:bic_likelihood_threshold} is larger.

\subsection{Representative fitted spectra and velocity resolution}
\begin{figure}[t]
\centering
\includegraphics[width=0.98\textwidth]{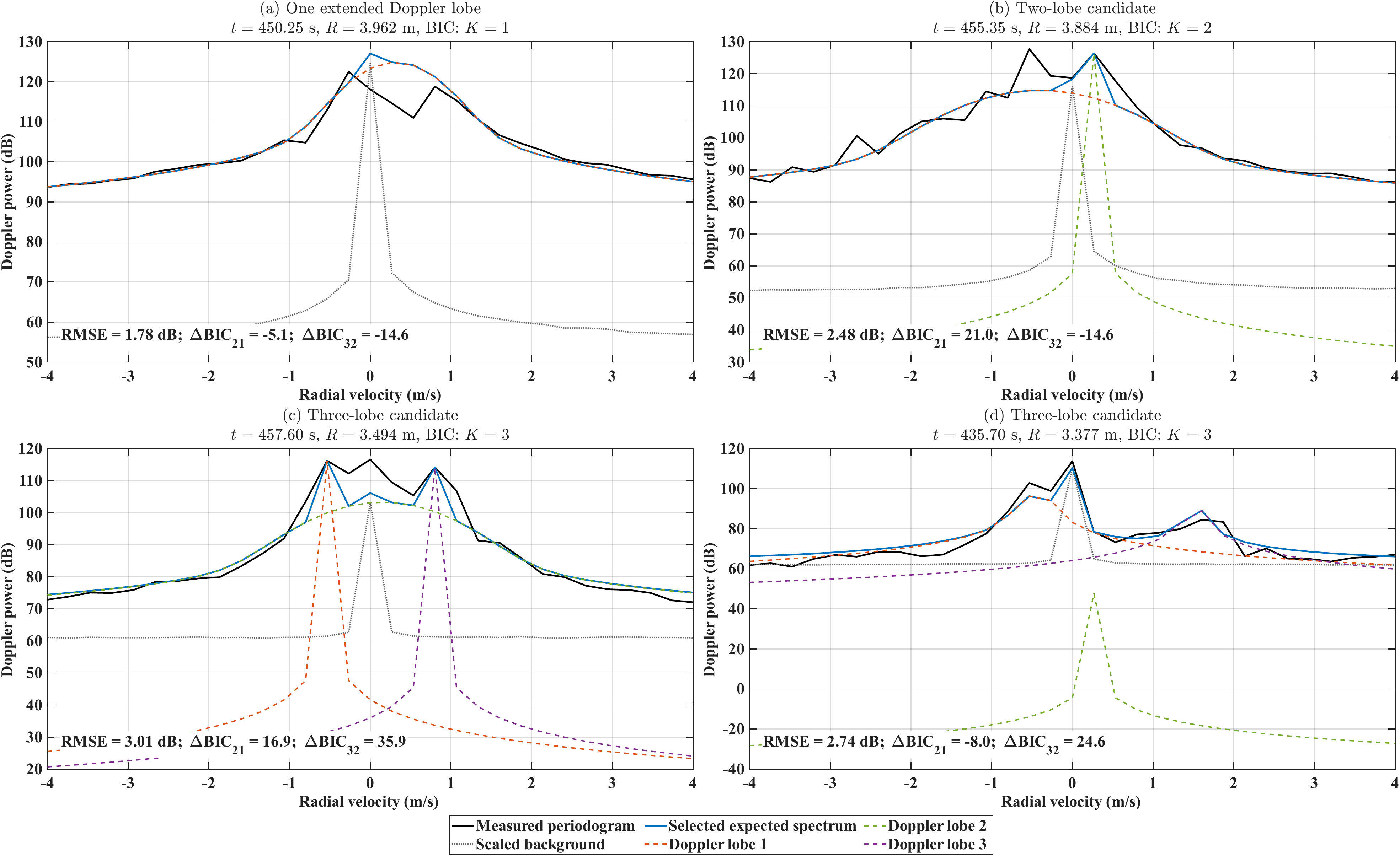}
\caption{Representative measured Doppler periodograms and scaled-background Whittle fits. Black curves denote the measurement, gray dotted curves denote the range-dependent background after multiplication by its fitted cell-dependent scale, blue curves denote the selected complete model, and dashed curves denote its individual extended Doppler-lobe contributions. All 128 non-oversampled ordinates enter the fit; the display is limited to \SIrange{-4}{4}{m/s} to resolve the lobe structure. Positive $\Delta\mathrm{BIC}_{21}$ or $\Delta\mathrm{BIC}_{32}$ means that the higher-order spectrum supplies sufficient likelihood improvement for the corresponding three additional parameters.}
\label{fig:fit_examples}
\end{figure}

Fig.~\ref{fig:fit_examples} compares representative measured periodograms with the selected expected spectra. The examples show how the information criteria react to spectral shape after the scaled no-motion background has been included in the model. Panel (a) is represented by one extended Doppler lobe with mean \SI{0.314}{m/s} and width \SI{0.353}{m/s}. Panel (b) supports two centres at \SI{-0.398}{m/s} and \SI{0.267}{m/s}, with $\Delta\mathrm{BIC}_{21}=21.0$. The opposite signs are consistent with simultaneous contributions towards and away from the radar within one range cell, but one fitted width is below the independent velocity resolution and is therefore unresolved.

Panels (c) and (d) show selected three-lobe cases. They illustrate the role of BIC as a spectral-shape diagnostic rather than a direct particle-population counter. Several fitted widths approach the numerical narrow-lobe limit, and the measured spectra remain jagged because each periodogram comes from one finite realization of a stochastic echo. These cases therefore support additional velocity centres in the fitted expected spectrum, while the precise narrow widths are not interpreted as resolved material dispersions. The unrestricted log-width parameterization prevents a hard lower bound at one Doppler bin; the resolution statement prevents over-interpretation of the resulting narrow values.

\section{Discussion}\label{sec:discussion}
The measurements establish that a compact \SI{60}{GHz} radar can look into {an} opaque bed and recover range-resolved spectral information about ensemble particle motion. The microscopic field is unpredictable because millimetric grain displacements change phase, grains cross the range-cell boundaries, and dense-bed propagation changes with the instantaneous arrangement. Its covariance and spectrum are nevertheless structured by the ensemble motion. The stochastic description is therefore a physical measurement model, not an attempt to resolve the motion of each and every particle. In applications like the one presented in this paper, ensemble motion carries more useful information than a device that resolves the motion of each and every particle. {For example, an important ensemble quantity in a fluidized bed is the variance of particle velocities about their mean, which contributes to the granular temperature.}

Pressure describes the mechanical condition of the complete bed, whereas the radar provides total power, mean radial velocity, and Doppler width as functions of range. Total radar power follows the imposed pressure evolution most clearly. The mean local width has only a weak positive association with pressure because it describes the velocity spread within a range cell rather than the overall bed condition. The correlations show that the radar response changes with the known operating sequence. 

The signed $K=1$ retrieval provides the clearest evidence of internal bidirectional motion. Positive and negative means coexist at different ranges and exchange their contribution to the total detected extended-lobe power even within one pressure condition. These quantities describe the dominant radial motion of the ensemble in each range cell. They do not provide individual particle trajectories or the complete {three-dimensional particle-velocity field}. Nevertheless, their spatial continuity, rapid reversals, and accompanying Doppler width provide radar evidence of irregular bidirectional motion consistent with bubbling-induced convective circulation.

The processing keeps the physical observation of bidirectional motion separate from spectral multiplicity. The no-motion records define a range-dependent background shape, while a positive scale adapts its strength to each later cell. A $K=1$ extended Doppler-lobe spectrum is fitted everywhere, and its integrated contrast with the scaled background determines whether the measured spectrum departs reproducibly from the no-motion calibration. Only this detected set enters the comparison of $K=1$, 2, and 3. The directional $K=1$ observation is therefore not conditioned on selecting a multi-lobe spectrum. Because the single-channel measurement cannot identify the physical origin of every persistent range-localized response, the motion interpretation additionally requires temporal evolution and continuity over neighbouring ranges.

The 128-point rectangular discrete Fourier transform (DFT) used for estimation provides an independent Doppler-bin spacing of \SI{0.2671}{m/s}. Zero padding creates a smoother displayed spectrum but no additional independent information and is therefore not used by AIC or BIC. A fitted width below this spacing indicates a contribution that is narrow relative to the available coherent record, but its numerical value cannot be interpreted as a precise particle-velocity dispersion. This distinction explains why evidence for two separated means is considerably more common than evidence for two resolved widths.

The localized $K=2$ fits strengthen the physical interpretation where two contributions are distinguishable inside one range-time cell. Their sparsity does not contradict the widespread positive and negative $K=1$ means: the latter occur across different ranges and times, while $K=2$ requires simultaneous spectral separation within one cell. The rare $K=3$ selections primarily expose the present resolution limit rather than establish three persistent particle populations.

The three one-lobe Doppler quantities also provide a practical route towards real-time observation. Each local spectrum is reduced to a small set of physically interpretable values at the \SI{20}{Hz} recorded update rate. A future implementation can therefore follow where the moving layer appears, whether the dominant motion reverses, and how the local {particle-velocity spread} changes without retaining every spectral ordinate. The uncommon multi-lobe fits add a targeted indication of cells in which simultaneous velocity contributions may be separable. The present work establishes the signal model and physical validation required for such monitoring; closed-loop control and quantitative hydrodynamic inversion require further calibration and independent local measurements.

The \SI{60}{GHz} system complements the higher-frequency measurements {discussed} in \cite{Bonmann2022,GuioPerez2023,Bryllert2023}. It sacrifices part of their range resolution but offers a different penetration-resolution compromise for a dense, solids-rich bed. The delayed response beyond the direct bottom interval also demonstrates why radar range in this environment must be interpreted as propagation delay rather than an unconditional geometric coordinate. Restricting the retrieval to the verified vessel interval prevents weak delayed paths from being interpreted as scattering from geometrically impossible locations below the vessel bottom. Establishing reliable range-resolved spectra, bidirectional motion measurements, a background-calibrated Doppler-spectrum detectability decision, and the limits of spectral separation is a necessary step before concentration inversion, multi-beam observations, or quantitative range-angle-Doppler imaging.
\section{Conclusion}\label{sec:conclusion}
The usefulness of a compact commercial \SI{60}{GHz} radar for fluidized-bed measurements has been demonstrated. The radar looks into the opaque facility and converts the stochastic return from an ensemble of electrically small particles into physically interpretable, range-resolved Doppler spectra. Particle displacement, exchange between range cells, shadowing, and changing propagation paths make the individual slow-time samples random, while their covariance and spectrum preserve the statistical motion of the ensemble. The resulting total power, mean radial velocity, and spectral width reveal where moving particles are detectable and how their radial motion evolves through the bed.

The complete-record and focused retrievals contain connected positive- and negative-mean regions that change repeatedly in range and time. Their power-weighted velocities and directional fractions demonstrate irregular bidirectional ensemble motion during developed fluidization, while the pressure signal supplies the corresponding bulk operating condition. A scaled no-motion background and finite-record Whittle likelihood formulation separate the detectable Doppler spectrum from stationary propagation and instrumental structure. AIC and BIC then determine whether each detectable local spectrum is represented adequately by one extended Doppler lobe or supports additional velocity centres. In the focused interval, BIC selects one lobe in 96.64\% of detectable cells and localizes the uncommon spectra that support two or three centres. Resolution checks distinguish evidence for separated centres from the more demanding claim of independently resolved lobe widths.

The novelty is a range-resolved statistical Doppler measurement that reveals internal bidirectional particle motion in a dense bubbling fluidized bed and determines when the local spectrum supports more than one distinguishable velocity contribution. Association with the independently recorded pressure response verifies that the radar observables follow the imposed fluidization sequence while adding information unavailable from {pressure sensors} alone. At one retrieval every \SI{50}{ms}, the compact Doppler quantities provide the basis for future real-time observation of bed expansion, motion reversals, and changing velocity spread. Future work will use multiple receive beams for finer angular observation and investigate polarization diversity as a means to separate contributions associated with particle motion, concentration, shape, and size.
\section*{Acknowledgment}
The authors acknowledge EnergieNL and GIDARA Energy for co-funding through the ``Inzetproject PPS programmatoeslag 2020 en 2021'' under project number CHEMIE.PGT.2022.014. The authors thank ir. Pascal J. Aubry for his help in carrying out the experiments. 
\bibliographystyle{IEEEtran}
\bibliography{references_v2_JTP,references}
\end{document}